\documentclass[fleqn,usenatbib]{mnras}

\usepackage{newtxtext,newtxmath}

\usepackage[T1]{fontenc}

\DeclareRobustCommand{\VAN}[3]{#2}
\let\VANthebibliography\thebibliography
\def\thebibliography{\DeclareRobustCommand{\VAN}[3]{##3}\VANthebibliography}

\usepackage{graphicx}	
\usepackage{amsmath}	
\usepackage{mathtools}
\usepackage{rotating}

\title[The GALAH survey: Open Clusters]{The GALAH survey: tracing the Galactic disk with Open Clusters}

\author[L. Spina et al.]{
Lorenzo Spina,$^{1,2,3}$\thanks{E-mail: \texttt{spina.astro@gmail.com}}
Yuan-Sen Ting,$^{4,5,6,7}$
Gayandhi~M.~De~Silva,$^{8,9}$
Neige Frankel,$^{10}$
Sanjib~Sharma,$^{2,11}$\newauthor
Tristan~Cantat-Gaudin,$^{12}$
Meridith~Joyce,$^{2,7}$
Dennis~Stello,$^{2,13}$
Amanda~I.~Karakas,$^{1,2}$\newauthor
Martin~B.~Asplund,$^{14}$
Thomas~Nordlander,$^{2,7}$
Luca~Casagrande,$^{2,7}$
Valentina~{D'Orazi},$^{1,3}$\newauthor
Andrew~R.~Casey,$^{1,2}$
Peter~Cottrell,$^{15}$
Thor Tepper-Garc\'ia,$^{2,11,16}$
Martina Baratella,$^{3}$
Janez~Kos,$^{17}$\newauthor
Klemen, \v{C}otar,$^{17}$
Joss~Bland-Hawthorn,$^{2,11}$
Sven~Buder,$^{2,7}$
Ken~C.~Freeman,$^{2,7}$
Michael~R.~Hayden,$^{2,11}$\newauthor
Geraint~F.~Lewis,$^{11}$
Jane~Lin,$^{2,7}$
Karin~Lind,$^{18}$
Sarah~L.~Martell,$^{2,13}$
Katharine J. Schlesinger,$^{2,7}$\newauthor
Jeffrey~D.~Simpson,$^{2,13}$
Daniel~B.~Zucker,$^{8,19}$
and Toma\v{z}~Zwitter$^{17}$
\\ \\
$^{1}$School of Physics and Astronomy, Monash University, VIC 3800, Australia\\
$^{2}$ARC Centre of Excellence for All Sky Astrophysics in Three Dimensions (ASTRO-3D)\\
$^{3}$Istituto Nazionale di Astrofisica, Osservatorio Astronomico di Padova, vicolo dell'Osservatorio 5, 35122, Padova, Italy\\
$^{4}$Institute for Advanced Study, Princeton, NJ 08540, USA\\
$^{5}$Department of Astrophysical Sciences, Princeton University, Princeton, NJ 08544, USA\\
$^{6}$Observatories of the Carnegie Institution of Washington, 813 Santa Barbara Street, Pasadena, CA 91101, USA\\
$^{7}$Research School of Astronomy \& Astrophysics, Australian National University, ACT 2611, Australia\\
$^{8}$Australian Astronomical Optics, Faculty of Science and Engineering, Macquarie University, Macquarie Park, NSW 2113, Australia\\
$^{9}$Macquarie University Research Centre for Astronomy, Astrophysics \& Astrophotonics, Sydney, NSW 2109, Australia\\
$^{10}$Max Planck Institute for Astronomy, K\"onigstuhl 17, D-69117 Heidelberg, Germany\\
$^{11}$Sydney Institute for Astronomy, School of Physics, A28, The University of Sydney, NSW 2006, Australia\\
$^{12}$Institut de Ci\`encies del Cosmos, Universitat de Barcelona (IEEC-UB), Mart\'i i Franqu\`es 1, E-08028 Barcelona, Spain\\
$^{13}$School of Physics, UNSW, Sydney, NSW 2052, Australia\\
$^{14}$Max Planck Institute for Astrophysics, Karl-Schwarzschild-Str. 1, D- 85741 Garching, Germany\\
$^{15}$School of Physical and Chemical Sciences, University of Canterbury, New Zealand\\
$^{16}$Centre for Integrated Sustainability Analysis, The University of Sydney\\
$^{17}$Faculty of Mathematics and Physics, University of Ljubljana, Jadranska 19, 1000 Ljubljana, Slovenia\\
$^{18}$Department of Astronomy, Stockholm University, AlbaNova University Centre, SE-106 91 Stockholm, Sweden\\
$^{19}$Department of Physics and Astronomy, Macquarie University, Sydney, NSW 2109, Australia\\
}

\date{Accepted, 12 January 2021 ---. Received, 28 October 2020}

\pubyear{2020}

\begin{document}
\label{firstpage}
\pagerange{\pageref{firstpage}--\pageref{lastpage}}
\maketitle

\begin{abstract}
Open clusters are unique tracers of the history of our own Galaxy's disk. According to our membership analysis based on \textit{Gaia} astrometry, out of the 226 potential clusters falling in the footprint of GALAH or APOGEE, we find that 205 have secure members that were observed by at least one of the survey. Furthermore, members of 134 clusters have high-quality spectroscopic data that we use to determine their chemical composition. We leverage this information to study the chemical distribution throughout the Galactic disk of 21 elements, from C to Eu. The radial metallicity gradient obtained from our analysis is $-$0.076$\pm$0.009 dex kpc$^{-1}$, which is in agreement with previous works based on smaller samples. Furthermore, the gradient in the [Fe/H] - guiding radius (r$_{\rm guid}$) plane is $-$0.073$\pm$0.008 dex kpc$^{-1}$. We show consistently that open clusters trace the distribution of chemical elements throughout the Galactic disk differently than field stars. In particular, at given radius, open clusters show an age-metallicity relation that has less scatter than field stars. As such scatter is often interpreted as an effect of radial migration, we suggest that these differences are due to the physical selection effect imposed by our Galaxy: clusters that would have migrated significantly also had higher chances to get destroyed. Finally, our results reveal trends in the [X/Fe]$-$r$_{\rm guid}$$-$age space, which are important to understand production rates of different elements as a function of space and time.
\end{abstract}




\section{Introduction}
\label{Introduction}

How is the Milky Way disk structured? How does it evolve with time? Where are stars primarily born in our Galaxy? After their formation, how do stars migrate within the Milky Way? Is it possible to trace the stars back to the location where they have formed? How are chemical elements synthesised in stars? Open stellar clusters are unique tools to answer these questions (e.g., see \citealt{Krumholz19} and references therein).

An open cluster is a group of stars that formed together, at the same time, from the same material and therefore have similar ages, distances from the Sun, kinematics and chemical compositions. As simple stellar populations, its members can be identified via kinematic, photometric, and spatial criteria (e.g., \citealt{Dias05,Kharchenko13,Cantat-Gaudin18,Liu19,Castro-Ginard19,Castro-Ginard20}). Furthermore, its age, mass and distance can be estimated in a relatively simple way through photometry (e.g., \citealt{Cantat-Gaudin18a,Cantat-Gaudin20,Monteiro19,Bossini19}). Open clusters are spread throughout the entire Milky Way disk and they span large ranges in ages (from $<$100 Myr to up to $\sim$8 Gyr). These are the unique properties that make open clusters ideal tracers of Galactic evolution in both space and time.

Among the several features, the distribution of chemical elements across the Galactic disk historically constitutes the most important constraint to chemo-dynamical models of our Milky Way. A number of studies (e.g., \citealt{Tosi88,Hayden14,Hayden15,Anders17}) have shown the spatial distributions of chemical abundances and their ratios across the Galactic disk. However, these studies are mainly based on field stars, which also include very old populations that had time to migrate significantly and redistribute the chemical elements across the Galaxy (e.g., \citealt{Sellwood02,Roskar12,Martinez-Medina16}). Open clusters are a valuable alternative, being on average younger \citep{Magrini17}, and therefore a better tracer of the gradients in the disk out of which the most recent stars formed. Since the work of \citet{Janes79}, much observational evidence has established that the metallicity distribution (often abbreviated by the iron-to-hydrogen ratio [Fe/H]) traced by clusters throughout the Milky Way disk shows a significant decrease with increasing distance from the Galactic centre. This ``radial metallicity gradient'' - in its apparent simplicity - reflects a complex interplay between several processes that are driving the evolution of our Galaxy, including star formation, stellar evolution, stellar migration, gas flows, and cluster disruption \citep{Cunha92,Cunha94,Friel95,Stahler04,Carraro06,Boesgaard09,Magrini09,Frinchaboy13,Netopil16,Anders17,Spina17,BertelliMotta18,Quillen18}. Complementary to the study of the overall metallicity distribution, the abundance ratios of several other elements, such as $\alpha$-elements, iron-peak, odd-Z, and neutron-capture, can provide deep insights on the variety of nucleosynthesis processes, with their production sites and timescales (e.g., \citealt{Carrera11,Ting12b,Reddy16,Duffau17,Magrini17,Magrini18,Donor20,Casamiquela20}). Therefore, understanding the distribution of metals traced by clusters across the Galactic disk is fundamental for explaining the birth, life and death of both stars and clusters, the recent evolution of our own Milky Way, and the evolution of other spiral galaxies \citep{Boissier00,Bresolin19}.

In this paper we leverage the datasets acquired by three Galactic surveys that are rapidly improving our knowledge of the Milky Way: APOGEE, \textit{Gaia}, and GALAH \citep{DeSilva15,Gaia16,Majewski17}. Astrometric solutions from \textit{Gaia} are used to identify the cluster stellar members observed by these surveys. Then, spectroscopic abundance determinations from APOGEE and GALAH for 134 open clusters have been used to trace the chemical distribution of elements throughout the Galactic disk. From this information we obtain critical insights into Galactic chemical evolution and on how our Galaxy shapes the cluster's demographic properties across the disk.

The manuscript is organised as follows: in Section~\ref{Datasets} we describe the dataset used in our analysis; the cluster membership analysis is detailed Section~\ref{Membership}; kinematic and chemical properties of open clusters are discussed in Sections~\ref{properties}; in Sections~\ref{gradient}, \ref{survivors}, and \ref{elements} we describe the chemical distribution of elements traced by the open cluster population and we discuss relevant implications; finally general conclusions are included in Section~\ref{conclusions}.

\section{Datasets}
\label{Datasets}

In this study we make use of three datasets: \textit{Gaia} DR2, APOGEE DR16, and GALAH+  \citep{GaiaCollaboration18,Ahumada20,Wittenmyer18,Sharma18,Sharma19,Buder20}.

Stellar coordinates and astrometric solutions from \textit{Gaia} are used for cluster membership analysis, which is described in Section~\ref{Membership}. Radial velocities provided by either \textit{Gaia}, APOGEE or GALAH are used in Section~\ref{Validation} to validate the membership analysis and in Section~\ref{kinematic} to determine the mean radial velocity of each cluster and its kinematic properties. Finally, chemical abundances from the two spectroscopic surveys are used to derive the chemical composition of open clusters in Section~\ref{Chem}.

\subsection{Stellar coordinates and astrometric solutions}
The ESA \textit{Gaia} space mission \citep{Gaia16} is providing an unprecedented three-dimensional map of our Galaxy in terms of number of observed targets, dimensionality, and astrometric precision and accuracy. In particular, the most recent Data Release, \textit{Gaia}
DR2 \citep{GaiaCollaboration18}, contains more than 1.6 billion sources as faint as G $\sim$21~mag. It provides coordinates ($\alpha$ and $\delta$), parallaxes ($\varpi$), proper motions ($\mu_\alpha$, $\mu_\delta$), magnitudes in three photometric bands (G, B$_{\rm p}$, and R$_{\rm p}$), and relative uncertainties, for more than 1.3 billion sources. This extraordinary dataset has become an invaluable resource in identifying Galactic open clusters and their stellar members (e.g., \citealt{Cantat-Gaudin18,Liu19,Castro-Ginard19,Castro-Ginard20}).

\subsection{Radial velocities}
\label{RVdatasets}
The GALAH+ and APOGEE DR16 datasets list stellar parameters (effective temperature T$_{\rm eff}$, surface gravity log~g, iron abundance [Fe/H], and microturbulence v$_{\rm mic}$) and/or radial velocity determinations for 600,258 and 473,307 stars, respectively. Additionally, \textit{Gaia} DR2 contains median radial velocities for 7,224,631 stars, with T$_{\rm eff}$ in the range 3550-6900 K. Radial velocity (RV) estimations from GALAH are used only for stars with $\tt{flag\_sp}$ = 0, $T_{\rm eff}\leq$8000 K and $\Delta$RV$<$3 km/s. Similarly, RV from APOGEE are used only for stars with $\tt{STARFLAG}$ $\ne$ [$\tt{BAD\_PIXELS}$, $\tt{SUSPECT\_RV\_COMBINATION}$, $\tt{BAD\_RV\_COMBINATION}$], $\tt{ASPCAPFLAG}$ $\ne$ [$\tt{ROTATION\_BAD}$], $T_{\rm eff}\leq$8000 K and $\Delta$RV$<$3 km/s. When a RV estimation is not provided neither by GALAH nor APOGEE, but is included in the \textit{Gaia} DR2 sample, then this latter is used. No corrections are applied to the RV values, because no significant offsets are found between the three datasets.

\subsection{Chemical abundances}
\label{Chemical_abundances}

The elements that we consider in this study are O, Na, Mg, Al, Si, Ca, Ti, V, Cr, Mn, Fe, Co, Ni, Cu, Zn, Y, Zr, Ba, La, Nd, and Eu. Among these, abundances of Zn, Y, Zr, Ba, La, Nd, and Eu are detected only by the GALAH survey, while all the other elements have been detected by both spectroscopic surveys.

\begin{figure}
 \includegraphics[width=\columnwidth]{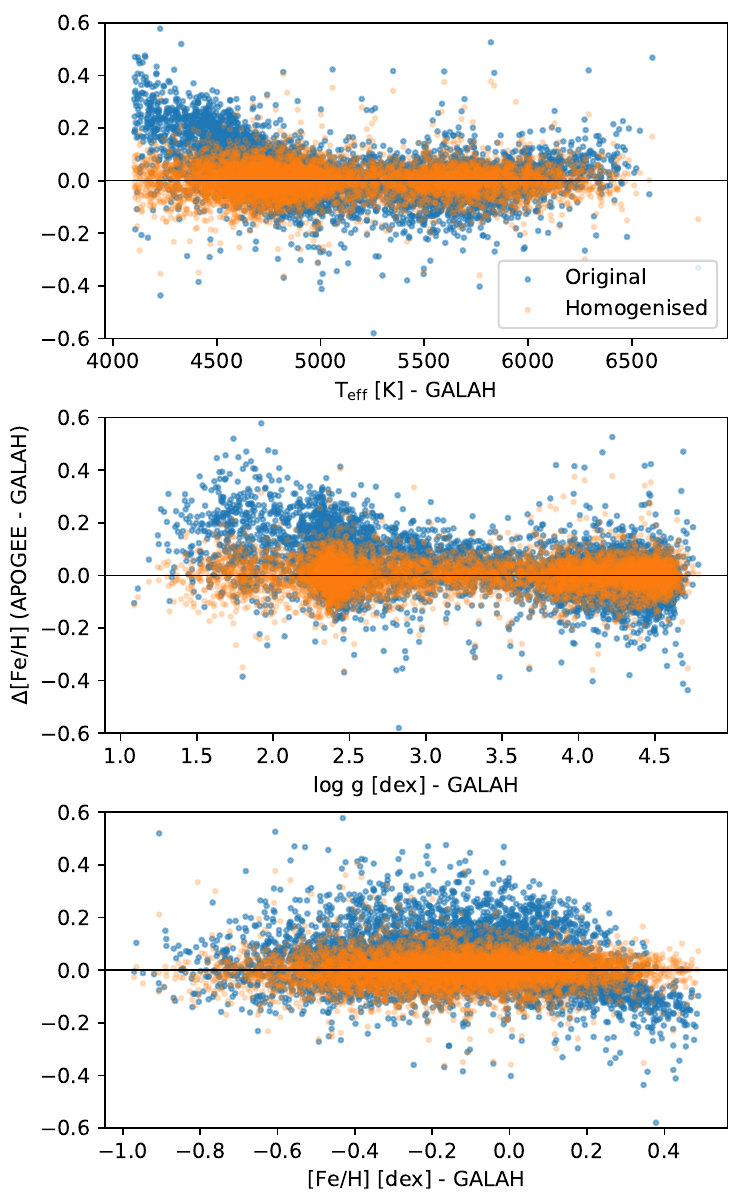}
 \caption{The panels show the differences in Fe abundances determined by the two surveys [Fe/H]$_{\rm APOGEE}$$-$[Fe/H]$_{\rm GALAH}$ as a function of the stellar parameters T$_{\rm eff}$, log~g, and [Fe/H] for a sample of stars that have been observed by both APOGEE and GALAH. The blue points represent the abundance differences before the homogenisation, while the orange points those of the homogenised dataset.}
 \label{homo_fe}
\end{figure}

The homogenising chemical abundances from two different datasets is typically a non-linear problem. In fact, abundance corrections for a give element may be a complex function of stellar parameters. This is especially true for the homogenisation of GALAH and APOGEE, as the two surveys utilise very different spectral ranges and linelists in their chemical analysis. As an example, Fig.~\ref{homo_fe} shows with blue dots the differential Fe abundances for a sample of stars observed by both GALAH and APOGEE as a function of the GALAH's stellar parameters.

The chemical abundances from the GALAH and APOGEE datasets are homogenised with a XGBoost\footnote{XGBoost (eXtreme Gradient Boosting) is a decision-tree-based ensemble Machine Learning algorithm that makes use of the gradient boosting technique to train regression or classification models. In few words, this algorithm produces a sequence of decision-tree models that are iteratively adjusted in order to improve their performances. For a more detailed description of XGBoost algorithms see \citealt{Chen16}.} Regressor. With this algorithm we train a separate model for the homogenisation of each element. A single model uses the stellar parameters T$_{\rm eff}$, log~g and [Fe/H] from GALAH as input features to predict the correction that needs to be applied to the chemical abundance of a specific element in order to homogenise the two datasets. More specifically, the target feature of a model is set to be the  difference between chemical abundances of a specific element given by GALAH and APOGEE  (e.g., [X/H]$_{\rm GALAH}$-[X/H]$_{\rm APOGEE}$). Models are trained over samples of stars that have been observed by both the spectroscopic surveys and that satisfy the following criteria: 4000$\leq \rm T_{\rm eff}\leq$7000 K, log g$\geq$0 dex, -1$\leq \rm [Fe/H]\leq$0.5 dex, v$_{\rm mic} \leq$2.5 km/s, $\Delta \rm T_{\rm eff}\leq$150 K,  $\Delta$log g$\leq$0.3 dex, and $\Delta$[Fe/H]$\leq$0.1 dex. Furthermore, to be part of the training sample for the homogenisation of a specific element, stars in the GALAH dataset must have $\tt{vbroad} \leq$60 km/s, $\tt{flag\_sp}$ = 0, and the element flag $\tt{flag\_elements\_fe}$ = 0. Similarly, the same star in the APOGEE dataset must have $\tt{VSCATTER}<$1 km/s, $\tt{STARFLAG}$ $\ne$ [$\tt{BAD\_PIXELS}$, $\tt{VERY\_BRIGHT\_NEIGHBOR}$, $\tt{LOW\_SNR}$], $\tt{ASPCAPFLAG}$ $\ne$ [$\tt{STAR\_BAD}$, $\tt{TEFF\_BAD}$, $\tt{LOGG\_BAD}$, $\tt{METALS\_BAD}$, $\tt{ALPHAFE\_BAD}$, $\tt{VMICRO\_BAD}$, $\tt{CHI2\_BAD}$, $\tt{ROTATION\_BAD}$], and the element flag $\tt{ELEMENT\_FE\_FLAG}$ = 0. Finally, the T$_{\rm eff}$ values determined by the two surveys have to be similar within 500 K, while the log~g values within 0.5~dex. These last selection criteria are on purpose very broad because our aim is to remove only the stars whose stellar parameters and abundances are affected by serious problematics in the analysis by one of the two surveys. This is because the dataset used to the model training should be similar, as much as possible, to the remaining dataset to which the model is applied. The sizes of these samples selected as above range from 2687 stars for V to 6569 stars for Fe. The 75$\%$ of these stars are used for the training samples, while the remaining 25$\%$ are part of the test samples.

A grid search is used to find the best parameters for the algorithm used to homogenise the Fe element. The best set of parameters is evaluated against the mean absolute error (MAE) and with a 5-fold cross-validation. The result of the homogenisation of the Fe abundances are shown in Fig.~\ref{homo_fe} with orange dots. The MAE values resulting from the models and calculated over the test samples range from 0.047 dex for Fe to 0.26 dex for V, with a median value of 0.078 dex. 

When applying a model to homogenise the abundances from one dataset to the other, it is important to verify that the training sample covers the same extent in the space of atmospheric parameters of the stars whose abundance we are going to correct. Since the stars in the GALAH sample generally cover a larger range in T$_{\rm eff}$ than the APOGEE sample, we apply the correction predicted by the models to the APOGEE abundance. The only exception is Fe, for which the APOGEE sample covers a larger extent in parameter space than GALAH. 

\begin{figure}
 \includegraphics[width=\columnwidth]{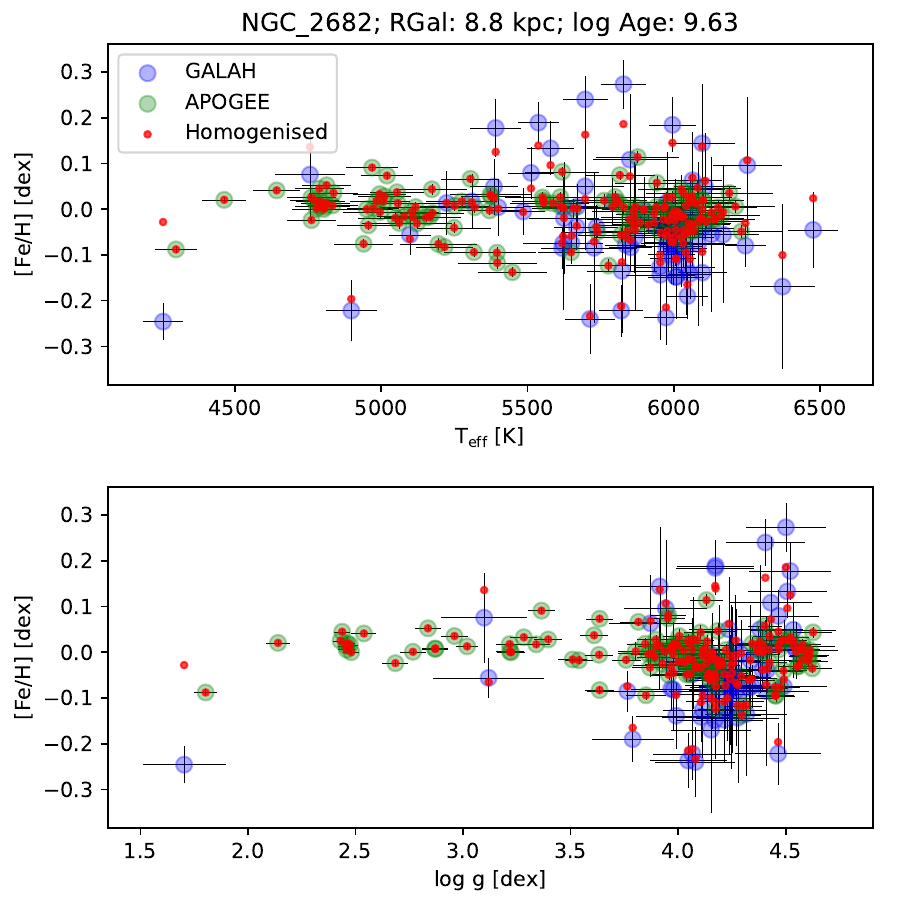}
 \caption{The figure shows the iron abundances of the stellar members of NGC~2682 (or M67) as a function of their T$_{\rm eff}$ (top panel) and log~g (bottom panel). The blue circles are the original abundances from GALAH, while the green circles are from APOGEE. The red dots represent the iron abundances for the same stars after the homogenisation. }
 \label{homo_check}
\end{figure}

Finally, the result of the entire homogenisation procedure is verified against cluster members: from a visual inspection of the chemical distribution of members of the same clusters observed by GALAH and APOGEE we exclude the presence of systematic shifts in the homogenised abundances or trends with atmospheric parameters (e.g., Fig.~\ref{homo_check}). This also implies that our following results are not affected by strong systematic effects in abundances between clusters due, for example, to different types offv stars that may be preferentially analysed in different clusters (e.g., differential 3D and NLTE effects or different lines in giants and dwarfs).

\section{Membership analysis}
\label{Membership}

The large astrometric and photometric survey performed by the \textit{Gaia} mission permits systematic and homogeneous studies of open clusters residing in our Galaxy. In this context, \citet{Cantat-Gaudin20} (hereafter CG20) published a catalog of more than 2000 Galactic open clusters, which also includes estimations of their fundamental properties, such as parallaxes ($\varpi$), proper motions ($\mu_\alpha$, $\mu_\delta$), ages, and the radius containing half the members (r$_{50}$), from which one can calculate $\sigma_{\rm dist}$ defined as r$_{50}$/0.6745. Along with this catalog, they also published a list of cluster members identified in the 3-dimensional astrometric space ($\varpi$, $\mu_\alpha$, $\mu_\delta$) through k-means clustering and with membership probability P$\geq$0.7, or with P=1 for clusters named UBC newly discovered by \citep{Castro-Ginard19,Castro-Ginard20}. Although this catalog is an invaluable resource for the study of open cluster demographics, a number of true cluster members may be missing either because previous membership analyses only considered stars with G$<$18 mag and within a restricted field of view or because membership probability was estimated to be lower than the applied threshold (i.e., 1 for all the UBC clusters and 0.7 for all the others).

Therefore, the first step of our study focuses on improving - where possible - the existing catalog of cluster members for the stellar associations that have been observed by APOGEE or GALAH. Our new membership analysis uses a Support Vector Machine\footnote{Support Vector Machines \citep{Vapnik1998} belong to kernel-based techniques and represent some of the most used machine learning algorithms for classification. The objective of a support vector machine algorithm is to find a hyperplane in an N-dimensional space (where N is the number of features) that distinctly classifies the data points. More specifically, the algorithm finds the hyperplane that has the maximum distance between data points of both classes (e.g., cluster members and field stars). However, not all data are linearly separable. In this case, support vector machines use \textit{kernels functions} to transform the space of the training dataset onto a higher dimension space so as to make it easier to linearly divided the data with a hyperplane.} classifier trained in the 5-dimensional space ($\alpha$, $\delta$, $\varpi$, $\mu_\alpha$, $\mu_\delta$) over samples of stars including the cluster members listed in CG20 and fiducial non-members.  Then, the results of this analysis are validated using RV values. The membership analysis is performed only for the 226 clusters with at least one GALAH or APOGEE target located within 3$\sigma_{\rm dist}$ from its centre. Likewise, in this work we focus only on associations older than 10 Myr, whose stellar populations are less dispersed in the 5-dimensional space of parameters and whose chemical abundances are more reliably determined than for younger associations. CG20 could not estimate an age for the very reddened clusters Berkeley 43, NGC 7419, and Teutsch 7. For these three clusters, we used the values estimated by \citet{Kharchenko09}.

Below we detail the steps of the membership analysis.

\subsection{Selection of analysis sample}
\label{sample}
The stars used for the membership analysis are selected according to the following criteria:
\begin{itemize}
\item Stars must be within 3$\sigma_{\rm dist}$ from the cluster centre.
\item Stars must have parallaxes $\varpi$ within $\varpi_{\rm cl}$ $\pm$ 5$\times$max($\varpi_{\rm cl}$, 1)$\times\sigma_{\varpi_{\rm cl}}$, where $\varpi_{\rm cl}$ is the cluster parallax from CG20 and $\sigma_{\varpi_{\rm cl}}$ the relative standard deviation. In addition, for clusters with $\varpi_{\rm cl}>$2 mas, we also rejected stars with $\varpi<$1 mas. Similarly, for clusters with $\varpi_{\rm cl}>$10 mas, we also rejected stars with $\varpi<$2 mas. Finally we rejected all stars with parallax error $\Delta \varpi>$1 mas. 
\item Stars must have proper motions $\mu_\alpha$ and $\mu_\delta$ that are within $\mu_{\alpha, cl}$ $\pm$ 5$\times \sigma_{\mu _{\alpha, cl}}$ and $\mu_{\delta cl}$ $\pm$ 5$\times \sigma_{\mu _{\delta, cl}}$, respectively. In addition, stars must have errors in proper motions ($\Delta \mu_\alpha$, $\Delta \mu_\delta$) smaller than 2 mas yr$^{-1}$.
\end{itemize}

\subsection{Training and test samples}
The training and test samples are selected from the sample described above. They must be composed exclusively of ``fiducial'' members and non-members of the clusters, which are stars with either very high or very low probability of being cluster members, based on our previous knowledge of the cluster stellar population. Namely, all the stars labelled as cluster members in CG20 are considered as fiducial members and are part of the training/test samples. 

The fiducial non-members, however, are defined as stars with a membership probability $<$0.02. This probability is calculated from the 5-dimensional multivariate normal distribution defined by the cluster mean parameters ($\alpha$, $\delta$, $\varpi$, $\mu_\alpha$, $\mu_\delta$) and the associated standard deviations from CG20. Generally, 10$\%$ of fiducial non-members are selected to be part of the training/test samples. However, in order to maintain a fair balance between the number of members and non-members, the number of non-members is forced to be between 0.2 and 2 times the number of members. Since the number of stars increases proportional to $\varpi^{-3}$, the selection of non-members is performed randomly within 5 different parallax bins equally spaced, in order to uniformly sample different distances from the Sun.

Finally the 75$\%$ of fiducial members and non-members is assigned to the training sample, while the remaining 25$\%$ forms the test sample.

\subsection{Model fitting}
A Support Vector Classifier from the $\tt{scikit-learn}$ Python library is trained using the sample described above. The five input features of the algorithm are $\alpha$, $\delta$, $\varpi$, $\mu_\alpha$, $\mu_\delta$, while the target parameter is whether the star is a cluster member or not. The Support Vector Classifier hyper-parameters C and $\gamma$ are chosen through a grid search with a 5-fold cross-validation that minimises the accuracy score. For the analysis, we adopt a Radial Basis Function kernel defined as k(X$_{1}$,X$_{2}$)=exp($-$$\gamma$d$_{1,2}$), where d$_{1,2}$ is the Euclidean distance between two points X$_{1}$ and X$_{2}$ of the train dataset.

The test sample is used to check the performance of the fitted model. The accuracy obtained from the test sample range from 0.8 to 1 for all the clusters, with a median value of 0.94.

\subsection{Membership probability}
The trained Support Vector Classifier model is applied to fit the entire sample selected as described in Section~\ref{sample}. Furthermore, we allow the algorithm to compute the cluster membership probability estimates. The uncertainties in the 5 features $\alpha$, $\delta$, $\varpi$, $\mu_\alpha$, $\mu_\delta$ and their correlations are taken into account by computing the membership probabilities 1000 times for each star. In each iteration the probability is calculated from a set of parameters randomly drawn from the 5-dimension multivariate normal distribution build using the astrometric solution, errors and correlation coefficients from \textit{Gaia} DR2. The final membership probability is the median of the 1000 probabilities.

The membership analysis is performed for the 226 open clusters older than 10~Myr and that potentially fall in the footprint of APOGEE or GALAH (as detailed in the second paragraph of this Section). The results of the membership analysis for these clusters are reported in Table~\ref{members}, which lists for every star with P$>$0 the \textit{Gaia} $\tt{source\_id}$, the probabilities resulting from our analysis and those from CG20. Finally, when available, the Table also lists RV values, the identification numbers from APOGEE and GALAH (i.e., $\tt{star\_id}$ and $\tt{APOGEE\_ID}$), atmospheric parameters, and homogenised abundances.

\subsection{Validation}
\label{Validation}

\begin{figure}
 \includegraphics[width=\columnwidth]{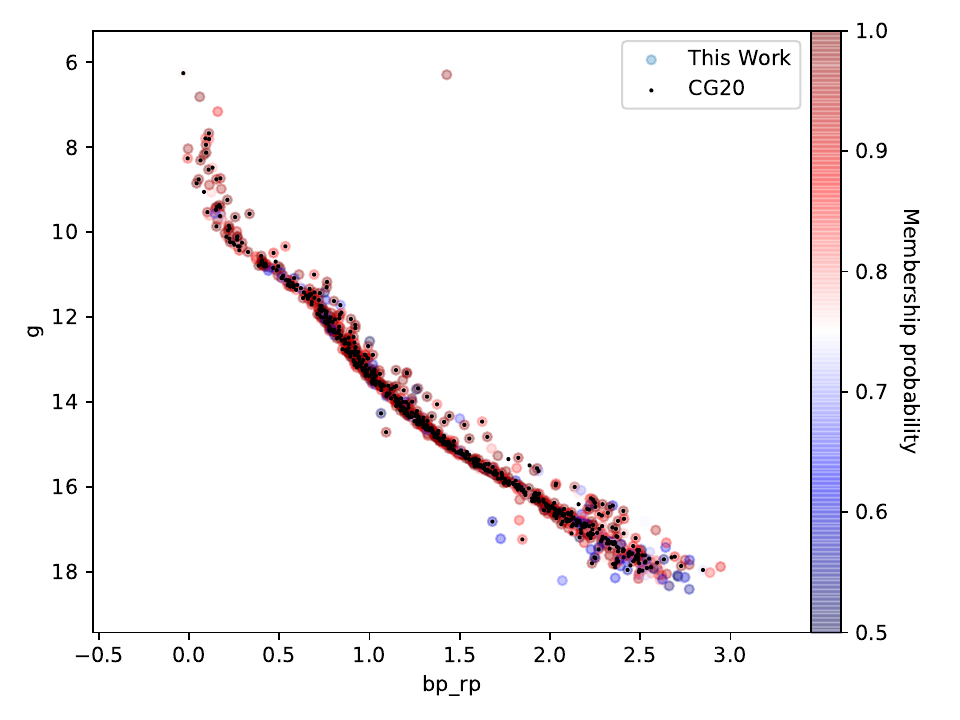}
 \caption{CMD of NGC 2516. Cluster members identified through our analysis are represented as circles colour-coded as a function of their membership probability. The smaller black dots represent the members listed in CG20.}
 \label{CMD}
\end{figure}

A first validation of the membership analysis consists of a visual inspection of the distribution traced in the colour-magnitude diagrams (CMD) by stars that have been labelled as cluster members (i.e., those with membership probability P$>$0.5). An example of these diagrams is shown in Fig.~\ref{CMD}. The cluster members with the highest probability are nicely aligned along the main sequence. The stars laying slightly above the main sequence are likely binary systems. A few stars are also located below the main sequence and are likely contaminants that are erroneously classified as members. Most of the members previously identified by CG20 are labelled as cluster members also by our analysis. However, there is no strict mutual correspondence between the classifications resulting from the two membership analyses.

Furthermore, we quantitatively assess the quality of our membership analysis by studying the RV distribution of cluster members. To do so, we use the RV values from APOGEE, GALAH, and \textit{Gaia}, described in Section~\ref{RVdatasets}. For each cluster we calculate two standard errors of the RV distributions traced by i) the members selected through our analysis (s.e.$_{\rm TW}$) and ii) the members from CG20 (s.e.$_{\rm CG20}$). We remind the reader that the standard error s.e.$\equiv\sigma$/$\sqrt{N}$, where $\sigma$ is the standard deviation and N is the size of the population. Therefore, while the standard deviation measures the spread of a population, the standard error indicates how accurately the population mean represents sample data and for this reason decreases with an increasing population size. The two membership analyses are then compared by dividing the standard error obtained from our membership to the one derived from the members listed in CG20: r=s.e.$_{\rm TW}$/s.e.$_{\rm CG20}$. Here we consider only the 65 clusters older than 100 Myr\footnote{Clusters older than 100 Myr are unlikely part of wide stellar associations that may have complex RV distributions.} and with at least three members from both our membership analysis and CG20. The mean value of the r ratios obtained for these clusters is equal to 0.99, which indicates that the two membership analyses have a nearly identical accuracy. However, our membership list provides an increase of the number of members with known RV by 6$\%$ relative to CG20.

\section{Kinematic and chemical properties of Open Clusters}
\label{properties}

Among the 226 clusters studied in Section~\ref{Membership}, 205 have candidate members that are observed by at least one of the two spectroscopic surveys and that have RV estimations. Hereafter, we consider the set of candidate members to encompass all the stars that have been labelled as members either through our membership analysis (i.e., P$>$0.5) or from the analysis performed by CG20. One of the key goals of our work is to use this information to determine the clusters' kinematic and chemical properties. 

To do so, we calculate the coordinates ($\alpha_{\rm cl}$, $\delta_{\rm cl}$), parallax ($\varpi_{\rm cl}$), proper motions ($\mu_{\alpha, cl}$, $\mu_{\delta, cl}$), and distance from the Sun (d$_{cl}$) and their uncertainties for each cluster. These are the error-weighted mean and standard deviation values obtained over the sample of candidate members. Note that physical distances from the Sun for every single member are obtained from \textit{Gaia} parallaxes and their uncertainties processed by $\tt{abj2016}$\footnote{Code available at \url{https://github.com/fjaellet/abj2016}.}, a Python module which implements the formalism from \citet{Astraatmadja16}. Also note that the estimation of cluster distances and the relative uncertainties assume the distance distributions of cluster members to be normal\footnote{We stress however, that given the excellent parallax quality and that the vast majority of our stars are closer than 5 kpc, this assumption only affects a very small number of clusters.}. The RVs from APOGEE, \textit{Gaia} and GALAH are then used to identify potential contaminants among the candidate members. Namely, for each cluster, we perform a 2-iteration 3-sigma clipping, using the RV median of the distribution as the centre value for the clipping. Stars that are rejected by the clipping are considered as contaminants and then removed from further analysis. All the other stars are considered as cluster members and are used to determine the median RVs of cluster populations and their standard deviations.

\subsection{Kinematic properties}
\label{kinematic}

The other kinematic properties of the open clusters are determined following the same procedure applied to GALAH stars, which is detailed in \citet{Buder20}, where the entire GALAH+ DR3 is described. Namely, using $\tt{GALPY}$\footnote{Code available at \url{http://github.com/jobovy/galpy}.} \citep{Bovy15} we transform the clusters' parameters obtained in Section~\ref{properties} (i.e., $\alpha_{\rm cl}$, $\delta_{\rm cl}$, $\varpi_{\rm cl}$, $\mu_{\alpha, cl}$, and $\mu_{\delta, cl}$) into Galactocentric coordinates and velocities, both cartesian (X, Y, Z, U, V, and W) and cylindrical (R$_{\rm Gal}$, $\phi$, z, v$_{\rm R}$, v$_{\rm \phi}$, v$_{\rm z}$). We also compute actions J$_{\rm R}$, L$_{\rm Z}$, J$_{\rm Z}$, guiding radii r$_{\rm guid}$, eccentricities e, and orbit boundary information (zmax, R$_{\rm peri}$, and R$_{\rm apo}$) in the Galactic potential $\tt{MWpotential2014}$ described in \citet{Bovy15} and a Staeckel fudge with 0.45 as focal length of the confocal coordinate system.

The statistical uncertainties of all these properties are obtained from a Monte Carlo simulation with a 10,000 sampling size. The final errors are equal to the standard deviations of the resulting distributions.

For the calculation, we adopt the distance of the Sun from the Galactic centre R$_{\odot}$~=~8.178 kpc \citep{Gravity18}, its height above the Galactic plane z$_{\odot}$~=~0.025 kpc \citep{Bennett19}, the Galactocentric velocity of the Sun (v$_{\odot, x}$, v$_{\odot, y}$, v$_{\odot, z}$)~=~(11.1, 12.24, 7.25) km s$^{-1}$ \citep{Schonrich10}, and a circular velocity of 229.0 km s$^{-1}$ \citep{Eilers19}.

All the values calculated here, including cluster coordinates, parallaxes physical distances, and proper motions are listed in Table~\ref{cluster_kin}.

\subsection{Chemical properties}
\label{Chem}

Among the 205 clusters with members observed by either one of the two surveys, 134 have stars with abundance determinations and that satisfy our selection criteria listed in Section~\ref{Chemical_abundances}. The homogenised abundances described in Section~\ref{Datasets} are used to calculate the chemical content of these open clusters.

Chemical abundances from spectroscopic surveys are obtained through automatic pipelines built to analyse, in a reasonable amount of time, thousands of spectra with very different signal-to-noise ratios and from stars having a broad range of atmospheric parameters, rotational velocities, and compositions. Therefore, due to the constantly accelerating acquisition of astronomical data by Galactic surveys, these pipelines often prioritise large statistics and rapid output over accuracy. Hence, in order to control the quality of the dataset used for our chemical analysis, the clusters' iron abundances [Fe/H] are derived from stellar members with $\Delta$T$_{\rm eff}<$150 K and spectra of $\tt{SNREV}\geq$100 (APOGEE) and $\tt{snr\_c2\_iraf}\geq$50 (GALAH). In addition, we apply a 3-sigma clipping before calculating the median [Fe/H] and standard deviation of each cluster population. 

Inaccurate abundance determinations especially affect elements other than iron, whose abundances are typically derived from a very limited number of absorption features. Furthermore, these inaccuracies often affect cool or hot stars, whose spectra are characterised by either blended or weak lines. For this reason, in order to calculate the median [X/Fe] of each open clusters, we only consider stellar members with T$_{\rm eff}$ between 4500 and 6500~K. In addition, since the open clusters' population is part of the \textit{thin disk}, their abundance ratios [X/Fe] must be well within $-$0.5 and $>$1 dex \citep{Donor20,Casamiquela20}. Therefore, in order to further limit the impact of false cluster members in our analysis, we reject all the values that are outside this range.

The chemical content of each open cluster is obtained from the median abundances of their members and the relative standard deviation. When a cluster has only one star with abundance determination, the standard deviation is replaced by the abundance uncertainty of that single star. The final chemical abundances [Fe/H] and [X/Fe] obtained for the 134 open clusters are listed in Table~\ref{cluster_chem}, together with the number of stars that are used for the analysis of each element.

Previous studies have used APOGEE and GALAH datasets to determine the chemical content of open clusters. For instance, \citet{Donor20} have determined the chemical composition of 126 open clusters exclusively from APOGEE data. Strikingly, only 76 of these are in common with our sample of 134 open clusters. This is because 31 clusters considered by \citet{Donor20} are not actually included in the CG20 catalog. Furthermore, the chemical content of other six clusters were based on stars that are not members of any association accordingly to both our membership analysis and the list of clusters' members in CG20. Finally, 13 clusters included in \citet{Donor20} are excluded by our analysis either because their stellar members do not satisfy the selection criteria listed in Section~\ref{Chemical_abundances} or because they are younger than 10 Myr. Another similar study by \citet{Carrera19} has determined the chemical composition of 90 open clusters using the APOGEE dataset. Among these, 69 are in common with our sample. The chemical content of the remaining 21 was based  on stars that either are not considered part of any association accordingly to our membership analysis or CG20 (i.e., 5 open clusters) or are excluded from our analysis because they do not satisfy our selection criteria (i.e., 16 clusters). \citet{Carrera19} also used the GALAH DR2 database \citep{Buder18} to determine the chemical composition of another 14 open clusters. All of those are included in our sample.

\section{The radial metallicity profile traced by Open Clusters}
\label{gradient}

In this Section we use the properties derived above to study the spatial distribution of metals across the Galactic disk traced by open clusters. Typically, the spatial distribution of metals is studied against the [Fe/H] - R$_{\rm Gal}$ diagram, which is described in Section~\ref{Fe-Rgal}. However, in Section~\ref{Fe-rguid}, we also discuss the distribution of metals as a function of the clusters' guiding radius r$_{\rm guid}$, which is defined as r$_{\rm guid}$= L$_{\rm z}$/v$_{\rm circ}$(r$_{\rm guid}$), where L$_{\rm z}$ is the angular momentum of the cluster and v$_{\rm circ}$(r$_{\rm guid}$) its circular velocity obtained from the Galactic potential. Since r$_{\rm guid}$ linearly scales with L$_{\rm z}$ and it provides additional important information that would otherwise be missing.

\subsection{The [Fe/H] - R$_{\rm Gal}$ diagram}
\label{Fe-Rgal}

\begin{figure*}
\includegraphics[width=\textwidth]{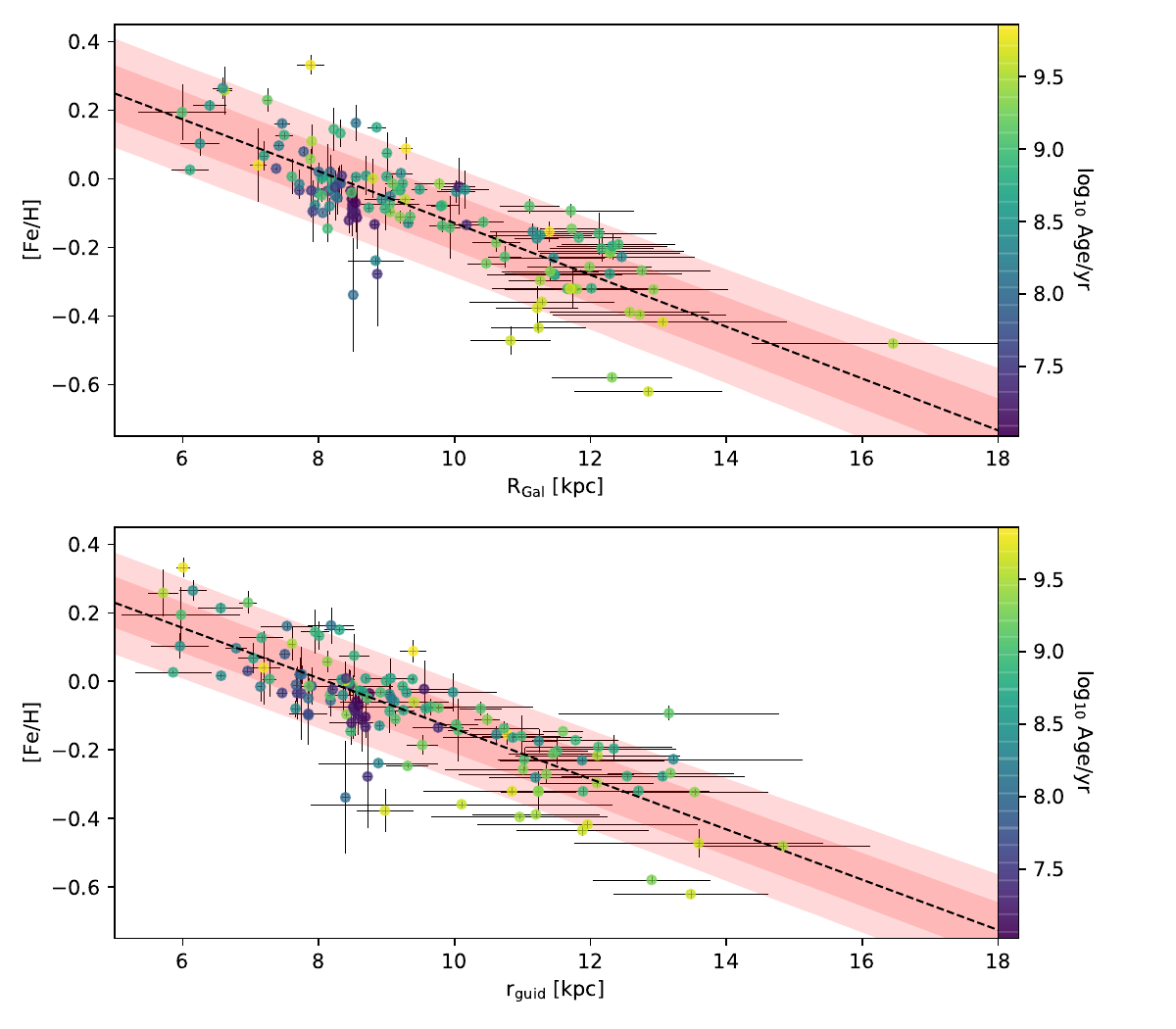}
 \caption{\textbf{Top.} The plot shows the cluster [Fe/H] values as a function of their Galactocentric distances R$_{\rm Gal}$. Clusters by circles colour coded as a function of the cluster age. Red shaded areas represent the 68 and 95$\%$ confidence intervals of the linear models resulting from the Bayesian regression, while the black dashed line traces the most probable model. \textbf{Bottom.} Same as in the top panel, but with [Fe/H] as a function of guiding radius r$_{\rm guid}$.}
 \label{FeH_dist}
\end{figure*}

In Fig.~\ref{FeH_dist}-top we plot the [Fe/H] abundance of 134 open clusters as a function of their distance from the Galactic centre R$_{\rm Gal}$. Symbols are colour coded as function of their ages. We model this distribution with a Bayesian regression using a simple linear model y$_i$ = $\alpha \times$ x$_i$ + $\beta$, where x$_i$ and y$_i$ are normal distributions centred on the R$_{\rm Gal}$ and [Fe/H] values of the $i^{\rm th}$ cluster:

\begin{equation}
\label{eq}
\begin{array}{l}
x_i = \mathcal{N}(\rm R_{\rm Gal, i},\sigma_{\rm R_{Gal}, i})\\
y_i = \mathcal{N}(\rm [Fe/H]_i,\sigma_{\rm Fe, i})
\end{array}
\end{equation}

The $\sigma_{\rm Fe}$ term in Eq.~\ref{eq} is the quadratic sum of the cluster abundance uncertainty $\Delta$[Fe/H] and a free parameter $\epsilon$ which accounts for the intrinsic chemical scatter between clusters at fixed Galactic radius. A variety of processes are responsible for this additional scatter that cannot be explained by measurement uncertainty, such as chemical evolution, radial migration and other unknown effects, including the possibility that uncertainties could be systematically underestimated. Priors for $\alpha$ and $\beta$ are chosen to be $\mathcal{N}$($-$0.068 dex/kpc, 0.1 dex/kpc) and $\mathcal{N}$(0.5 dex, 1 dex), respectively. Our prior for the $\epsilon$ parameter is a positive half-Cauchy distribution with $\gamma$=1. We run the simulation with 10,000 samples, half of which are used for burn-in, and a No-U-Turn Sampler \citep{Hoffman11}. The script is written in Python using the $\tt{pymc3}$ package \citep{Salvatier16}. 

The convergence of the Bayesian regression is checked against the traces of each parameter and their autocorrelation plots. The 68 and 95$\%$ confidence intervals of the models resulting from the posteriors are represented in Fig.~\ref{FeH_dist}-top with red shaded areas. The distribution of clusters in the diagram is nicely captured by the posteriors, with no need to apply a broken-line regression analysis. This is because the break in the radial metallicity profile is expected near the Galactocentric distance of the outermost group of clusters in our sample (i.e., R$_{\rm break}\sim$ 11-13 kpc) or beyond (e.g., \citealt{Yong12,Donor20}). 

Nevertheless, we identify some clusters that are clearly outliers in relation to the main distribution traced by the entire sample, such as the metal-rich NGC~6791 (R$_{\rm Gal}$ = 7.8 kpc; [Fe/H]=0.32 dex) and the metal-poor NGC~2243 (R$_{\rm Gal}$ = 10.9 kpc; [Fe/H]=$-$0.47 dex). These outliers are predominantly old clusters (age$>$1 Gyr), that have had the time to migrate significantly across the Galactic disk.

The results of the linear regression are also listed in Table~\ref{posteriors_tab}, which reports the mean, standard deviation, and 95$\%$ confidence interval of each posterior distribution. Interestingly, our radial metallicity gradient $\alpha$ = -0.076$\pm$0.009 dex kpc$^{-1}$ is consistent with the values given by recent studies based on open clusters, such as \citet{Jacobson16}, \citet{Carrera19} and \citet{Donor20}, who found gradients equal to $-$0.10$\pm$0.02, $-$0.077$\pm$0.007 and $-$0.068$\pm$0.004 dex kpc$^{-1}$, respectively. On the other hand, our slope is 1.8$\sigma$ steeper than the value obtained by \citet{Casamiquela19}, i.e., $-$0.056$\pm$0.011 dex kpc$^{-1}$. However, their linear regression was performed though a linear regression coupled with an outlier rejection \citep{Hogg10}.

Interestingly, between 7 and 9 kpc from the Galactic centre, where clusters have a very broad range of ages (i.e., from 10 Myr to 6.8 Gyr), the bulk of the youngest associations are the most metal-poor and they are located below the black dashed line, which represents the most probable model resulting from the Bayesian linear regression. On the other hand, clusters older than 1 Gyr have higher metallicities. Under the assumption that these clusters have not migrated significantly and that their metallicity represents the chemical content of the gas from which they have formed, our observation is at odds with chemical evolution models of our Galaxy (e.g., \citealt{Magrini09,Minchev13,Minchev14}), which predict an increase of the metallicity as time goes on. This apparent contradiction was previously noticed both in the solar neighbourhood \citep{James06,DOrazi09d,DOrazi09b,Biazzo11a,Spina14} and beyond \citep{Spina17}, where the youngest stellar associations all have similar metal content to one another regardless of their distance from the Galactic centre, and that they typically are more metal poor compared to older clusters located at similar Galactocentric radii. The same contradiction was also observed in much smaller-scale environments, such as the Orion complex (\citealt{Biazzo11a,Biazzo11b}; Kos, submitted), where the young Orion Nebula Cluster (2-3 Myr) is more metal poor than the older $\lambda$ Ori and 25 Ori subclusters (5-10 Myr). It is unlikely that these unexpected results, in evident contrast with predictions from Galactic chemical evolutionary models, are the consequence of a decreasing metal content in the interstellar medium. Instead they seem to be linked to systematics in the spectroscopic analysis due to stellar activity, which is typically neglected, but it is particularly strong in young stars \citep{Lorenzo-Oliveira18} and it can make stellar atmospheres appear to be more metal poor than what they actually are \citep{Galarza19,Baratella20,Spina20}. 

\setcounter{table}{3}
\begin{table}
\caption{Posteriors linear regression}
\centering
\label{posteriors_tab}
{\small
\begin{tabular}{cccc}
\hline
Parameter & Mean & $\sigma$ & 95$\%$ C.I.  \\ \hline \hline
\multicolumn{4}{|c|}{[Fe/H] - R$_{\rm Gal}$} \\ \hline
$\alpha$ [dex kpc$^{-1}$ ] & -0.076 & 0.009 & $-$0.085 - $-$0.066\\
$\beta$ [dex] & 0.63 & 0.09 & 0.54 - 0.72\\
$\epsilon$ [dex] & 0.083 & 0.013 & 0.069 - 0.096 \\ \hline \hline
\multicolumn{4}{|c|}{[Fe/H] - r$_{\rm guid}$} \\ \hline
$\alpha$ [dex kpc$^{-1}$ ] & -0.073 & 0.008 & $-$0.082 - $-$0.065\\
$\beta$ [dex] & 0.60 & 0.08 & 0.52 - 0.68\\
$\epsilon$ [dex] & 0.074 & 0.013 & 0.062 - 0.087\\ \hline
\end{tabular}
}
\end{table}

\subsection{The [Fe/H] - r$_{\rm guid}$ diagram}
\label{Fe-rguid}

It is well known that stars can travel a long way from the orbits in which they were born, which can complicate determinations of their origin. This stellar migration has shaped the distribution of elements that we observe today across the Galactic disk and represents one of the biggest limitations to our understanding of the chemical evolution of the Milky Way (e.g., \citealt{Roskar08,Schonrich09,Frankel20,Sharma20}). Following the terminology introduced by \citet{Sellwood02}, radial mixing of stars across the disk can be of two types: churning and blurring. Churning defines stars migrating due to a progressive gain or lose angular momentum, without significant change in the radial action J$_{\rm R}$, from resonant interactions with the potential of non-axisymmetric structures (such as spirals and bars). In contrast, blurring conserves the angular momentum of the individual star, but it heats the disk in all directions due the radial or vertical oscillations of stars with J$_{\rm R}$ or J$_{\rm Z}$ $>$ 0 kpc km/s. 

In exactly the same way, open clusters can also migrate across the disk through churning and blurring. This can explain the presence of several outliers in Fig.~\ref{FeH_dist}-top, such as NGC~6791 which probably was formed within the innermost regions of the Galaxy, where the interstellar medium is characterised by higher metal abundances, and then it migrated outward where it is observed today.

Both churning and blurring are frequently invoked to explain the degree of metallicity spread observed at a given Galactocentric distance for both open clusters and field stars (e.g., \citealt{Casagrande11,Ness16,Quillen18}). Precise knowledge of stellar orbital parameters in our Galaxy would ideally allow us to assess the relative importance of the two types of radial mixing (e.g., \citealt{Kordopatis15,Minchev18,Hayden18,Hayden20,Frankel20,Feltzing20}). This is a topic of much debate, as it is critical to our understanding of the mixing processes in action across the Galactic disk, with extremely important implications for models of evolutionary history of the Milky Way \citep{Haywood13,Spitoni19}.

A direct consequence of radial mixing is that the current Galactocentric distance of open clusters is not always representative of their birth location. This is especially true for old open clusters, as they have already completed several orbits around the Galactic centre. Their Galactocentric distance may have significantly changed with time either through blurring, churning or a combination of the two. As a result these two types of stellar migration, the intra-cluster chemical dispersion at each R$_{\rm Gal}$ value is expected to increase with time. However, since angular momentum primarily changes with churning, L$_{\rm z}$ is expected to be a more fundamental parameter than R$_{\rm Gal}$ against which the chemical distribution of elements can be studied. This is because the degree of intra-cluster chemical scatter as a function of angular momentum is primarily preserved by blurring and only increases with churning. Therefore, comparing the chemical dispersion of open clusters in the [Fe/H]-R$_{\rm Gal}$ and [Fe/H]-L$_{\rm z}$ spaces could allow us to assess whether churning or blurring is the dominant type of mixing or if cluster migration is affected by a balanced combination of the two.

In Fig.~\ref{FeH_dist}-bottom we show the cluster metallicities [Fe/H] as a function of their guiding radii r$_{\rm guid}$, defined as the radii of a circular orbits with angular momenta L$_{\rm z}$. Clusters such as NGC~6791 or NGC~2243, that in the [Fe/H]-R$_{\rm Gal}$ diagram show an extreme deviation from the other clusters (see Fig.~\ref{FeH_dist}-top) because they migrated during their lifetimes, are now more aligned to the main distribution in the [Fe/H]-r$_{\rm guid}$ diagram (Fig.~\ref{FeH_dist}-bottom). This suggests that cluster migration is not entirely due to churning, but that also blurring has an important role in it. On the other hand, clusters such as Berkeley~32 (r$_{\rm guid}$ = 9.0 kpc; R$_{\rm Gal}$ = 11.3 kpc; [Fe/H]=$-$0.36 dex) become strong outliers in the [Fe/H]-r$_{\rm guid}$ diagram, indicating that either the intrinsic chemical scatter cannot be totally explained by blurring alone or that clusters with similar birth location can have very different initial L$_{\rm z}$. However, this latter hypothesis seems less likely, as the location of the youngest clusters (e.g., log Age/yr $<$ 7.5) in Fig~\ref{FeH_dist} indicates that freshly formed stars with identical R$_{\rm Gal}$ also have very similar L$_{\rm z}$.

\begin{figure}
 \includegraphics[width=\columnwidth]{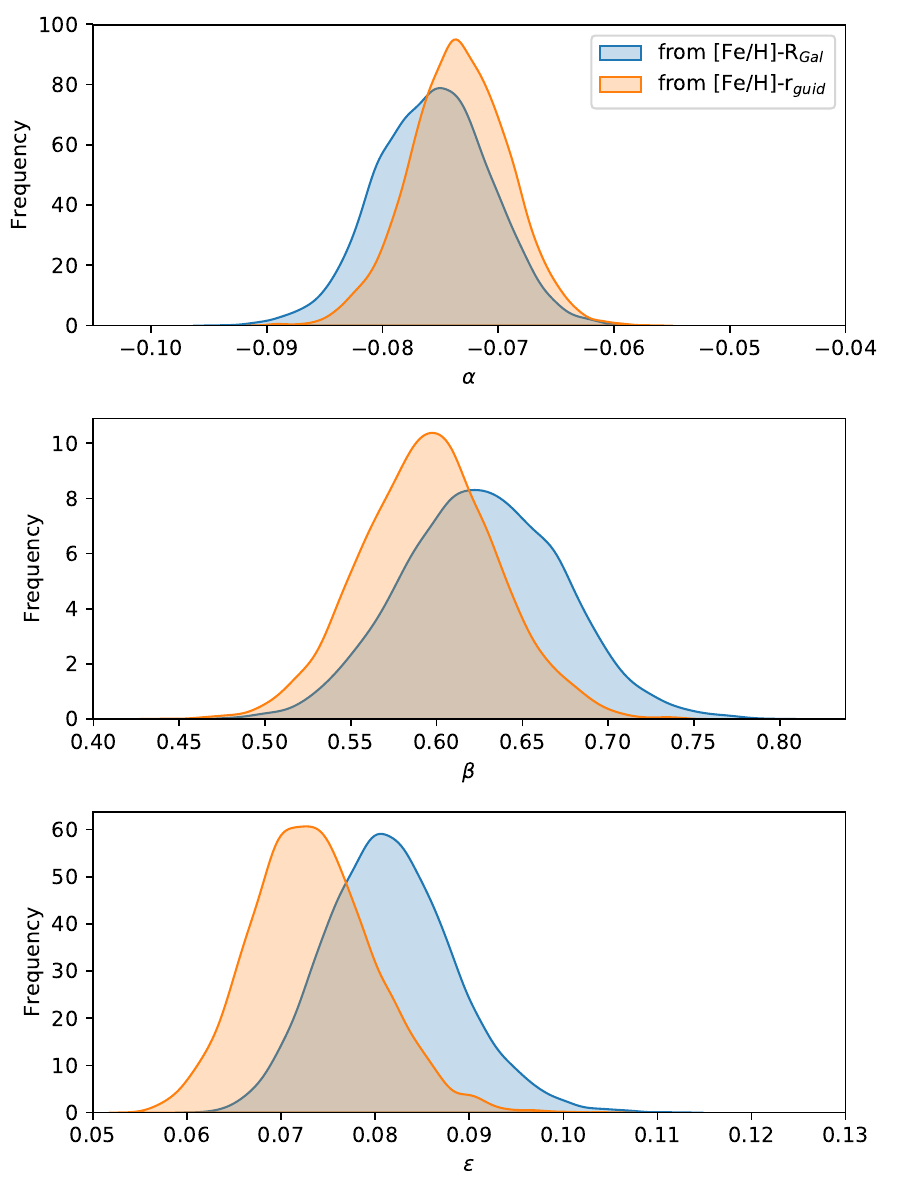}
 \caption{Posterior distributions for the parameters $\alpha$, $\beta$, and $\epsilon$ obtained from the Bayesian linear regression of the open clusters' distributions in the [Fe/H]-R$_{\rm Gal}$ and [Fe/H]-r$_{\rm guid}$ diagrams.}
 \label{all_posteriors}
\end{figure}

In order to perform a more quantitative comparison between the two diagrams in Fig~\ref{FeH_dist}, we perform a Bayesian regression of the cluster distribution in [Fe/H]-r$_{\rm guid}$ space, repeating the same procedure described in Section~\ref{Fe-Rgal} for the [Fe/H]-R$_{\rm Gal}$ diagram. The posterior distributions obtained from the two diagrams are shown in Fig.~\ref{all_posteriors} and their 95$\%$ confidence intervals are listed in Table~\ref{posteriors_tab}. The $\alpha$ and $\beta$ posteriors are identical in both diagrams. On the other hand, the $\epsilon$ posterior obtained from the [Fe/H]-r$_{\rm guid}$ diagram spans lower values than the one obtained from [Fe/H]-R$_{\rm Gal}$. This implies that the intrinsic chemical scatter, which is the scatter that cannot be explained by measurement uncertainties, is lower in the [Fe/H]-r$_{\rm guid}$ diagram than in [Fe/H]-R$_{\rm Gal}$. Note that the uncertainties on the y-axis are identical in both the diagrams and also those on the x-axis are very similar. In fact, the median and standard deviation of the differences between the two uncertainties (i.e., $\Delta$R$_{\rm Gal}$-$\Delta$r$_{\rm guid}$) are $-$0.04 and 0.26 kpc, respectively. Thus, the smaller intrinsic scatter that is characteristic of the [Fe/H]-r$_{\rm guid}$ diagram suggests that r$_{\rm guid}$ is a better approximation of the birth radius than R$_{\rm Gal}$. 

A first order quantification of the blurring contribution can be roughly estimated through the difference R$_{\rm Gal}$-r$_{\rm guid}$. The standard deviation of these differences calculated over the sample of 134 clusters shown in Fig.~\ref{FeH_dist} is equal to 0.64 kpc. Thus, assuming a radial metallicity profile described by the values in Table~\ref{posteriors_tab}, a blurring contribution of 0.64 kpc is responsible for an intrinsic chemical scatter of 0.04 dex. Interestingly, this latter value is equal to the quadratic difference between the $\epsilon$ posteriors obtained from the [Fe/H]-r$_{\rm guid}$ and the [Fe/H]-R$_{\rm Gal}$ diagrams. Thus, we conclude that blurring is responsible of the higher $\epsilon$ posteriors measured from the [Fe/H]-R$_{\rm Gal}$ diagram.

Although we are not able to measure the relative importance of blurring over churning, this result implies that churning is not the sole or dominant type of mixing process for open clusters. In fact, if churning was the sole reason for migration, all the kinematic information around the clusters' birth locations would have been completely erased from their angular momenta making the $\epsilon$ posteriors from the [Fe/H]-r$_{\rm guid}$ and [Fe/H]-R$_{\rm Gal}$ diagrams looking very similar (Fig~\ref{all_posteriors}). This conclusion is different to what is found for field stars, whose migration process is dominated by churning \citep{Frankel20,Sharma20}. 

The remaining intrinsic chemical scatter observed in the [Fe/H]-r$_{\rm guid}$ diagram could be attributable to a broad variety of factors, including i) churning \citep{Quillen18}; ii) a clumpy and transversely anisotropic distribution of metals within the disk or azimuthal metallicity variations \citep{Pedicelli09,Wenger19}; iii) stellar nucleosynthesis, which progressively produces new elements that are eventually released into the interstellar medium enriching our Galaxy of metals \citep{Kobayashi20}; iv) magnetic fields and chromospheric activity that make young stars appear more metal poor than what they actually are \citep{Spina20}; v) a higher-order polynomial radial metallicity distribution \citep{Maciel19}; vi) the warp of the Galactic disk, caused by the angular momentum vector of the outer disk which is not aligned with that of the inner disk - if neglected like in our analysis, the warp can lead to an underestimation of both r$_{\rm guid}$ and R$_{\rm Gal}$ for a fraction of the outermost clusters \citep{Amores17,Chen19} - and vii) eventual systematic errors in our analysis that could have led to the underestimation of the uncertainties.

\bigskip

\section{Survivors of the Galaxy}
\label{survivors}

An increasing number of studies have shown that radial mixing plays an important role in redistributing stars across the Milky Way, hence shaping their demographics throughout the entire Galaxy disk (e.g., \citealt{Frankel18,Hayden18,Hayden20,Feltzing20}). In Section~\ref{Fe-rguid} we have confirmed that open clusters are not exempt from this process, which - in fact - is expected to act in the same way for stars and clusters. Even though, we must ask whether it is possible that other processes are differentiating the way stars and clusters are re-distributed across the disk, or if instead we can indiscriminately apply the same chemo-dynamical models both to field stars and open clusters in our Galaxy. 

\begin{figure}
 \includegraphics[width=\columnwidth]{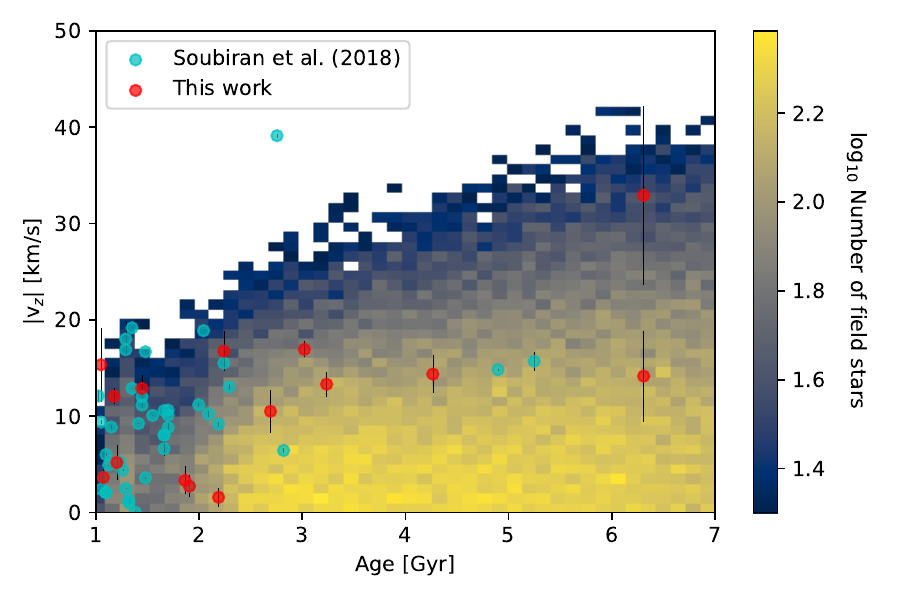}
 \caption{Absolute value of the vertical velocity |v$_{\rm z}$| as a function of age for open clusters in our sample (red dots), open clusters from \citet{Soubiran18} (cyan circles), and field stars (2D density histogram). The plot shows only clusters and stars located within 5 and 9 kpc from the Galactic centre and older than 1 Gyr. The field stars are those observed by GALAH.}
 \label{age_vz}
\end{figure}

In this regard, it is interesting to compare the demographics of open clusters living in the inner Galaxy, which is stormed by strong gravitational potentials (e.g., central bar, spirals, dense giant molecular clouds), to those of field stars located within the same region. In Fig.~\ref{age_vz} is shown the absolute value of the vertical velocity |v$_{\rm z}$| as a function of age for the clusters studied in our work (red circles), open clusters from \citet{Soubiran18} (cyan circles), and thin disk stars\footnote{We define thin disk stars as those with [$\alpha$/Fe] $<$ 0.1 dex.} observed by GALAH (density plot). These latter are i) main-sequence turn-off stars included in GALAH DR3, and ii) red giant stars with asteroseismic information from the NASA K2 mission and spectroscopic followup by the K2-HERMES survey \citep{Sharma19}. The ages of field stars are computed with the BSTEP code \citep{Sharma18}, which provides a Bayesian estimate of intrinsic stellar parameters from observed parameters by making use of stellar isochrones. In Fig.~\ref{age_vz} we only consider clusters and stars that are located within 5 and 9 kpc from the Galactic centre and older than 1 Gyr. As expected from previous studies (e.g., \citealt{Hayden17,Ting19}) the range of v$_{\rm z}$ values covered by field stars increases with age. However, the bulk of the stellar distribution is concentrated at low |v$_{\rm z}$| values, regardless of the age. Due to the lack of many data points, it is not clear whether the range of v$_{\rm z}$ values covered by open clusters increases with age or not. Instead, what it is evident from the plot is that open clusters, especially those older than $\sim$2 Gyr, are mostly absent for |v$_{\rm z}$| $\lesssim$ 10 km s$^{-1}$. The result does not change if we use the vertical action J$_{\rm Z}$ instead of |v$_{\rm z}$|. A similar conclusion was reached also by \citet{Soubiran18}. They noticed that the old population of open clusters is different to the younger one in terms of vertical velocity, with clusters older than 1 Gyr exhibiting a much wider range of v$_{\rm z}$ values than the younger clusters. These pieces of evidence strongly suggest that clusters living near the Galactic midplane undergo a quicker disruption than clusters living most of their lives far from it. 

The fact that the Galactic disk seems highly efficient in disrupting open clusters is a key factor in understanding how the Milky Way itself imposes biases between demographics of field stars and clusters and how they migrate across the disk. Namely, when non-axisymmetric potentials in our Galaxy kick a single star, the energy is fully converted to a variation in the stellar orbital action-angle coordinates, including its angular momentum (e.g., \citealt{Sellwood02,Roskar12,Martinez-Medina16,Mikkola20}). On the other hand, kicks given to an open cluster also contribute to a heating of the population of cluster members, with a number of stars that are stripped away from their native association at each encounter with these tidal forces (e.g., \citealt{Gieles07,Fujii12,Martinez-Barbosa16,Webb19,Jorgensen20}). After a certain number of disruptive interactions with these potentials, the cluster dissipates into field stars.

Therefore, the clusters that we observe today and that we use to trace the chemical distribution of elements across the Galactic disk (see Fig. \ref{FeH_dist}) are either those that are young enough to not have undergone numerous interactions with the gravitational potential or old clusters living most of their time far from these potentials (e.g., far from the mid plane or in the outer disk; see Fig.~\ref{age_vz}). Either way, these are the clusters that have probably conserved most of their angular momentum and for which we expect a smaller degree of churning than a typical star that has lived its entire life in the mid-plane, bouncing between potentials.

This Galactic selection effect is expected to have significant repercussions on the general demographics of existing open clusters compared to those of field stars and, in particular, on how these two populations trace the distribution of chemical elements across the disk. Below we discuss two pieces of observational evidence of this bias.

\subsection{The age dependence of the radial metallicity gradients}

Studying the age dependence of the radial metallicity gradient is informative about the production rate of metals at different locations in the Galaxy, but it also gives insights on how metals are subsequently carried across the disk due to radial mixing. Several works (e.g., \citealt{Friel02,Magrini09,Carrera11,Yong12,Cunha16,Spina17,Donor20}) have shown that the radial metallicity gradient traced by open clusters flattens with decreasing age (the literature is not, however, in uniform consensus; see \citealt{Salaris04})

\begin{figure}
 \includegraphics[width=\columnwidth]{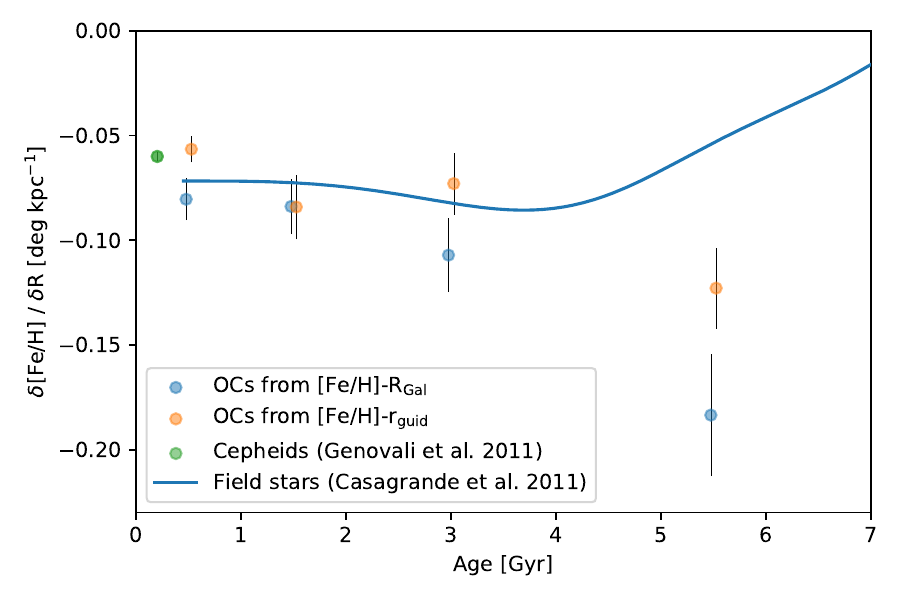}
 \caption{Age dependence of the Galactic metallicity gradient traced by open clusters in the [Fe/H]-R$_{\rm Gal}$ (blue dots) and [Fe/H]-r$_{\rm guid}$ (red dots) diagrams. The gradient age dependence traced by Cepheids \citep{Genovali14} is represented by a green circle, while field stars \citep{Casagrande11} are represented by a solid line.}
 \label{gradient_evolution}
\end{figure}

We split the sample of open clusters into four groups depending on their ages: $<$1 Gyr,  1$\leq$ age $<$2, 2$\leq$ age $<$4, and $\geq$4 Gyr. These bins are arbitrarily chosen to contain a balanced numbers of clusters, which is equal or larger than 10 clusters in every group. Fig.~\ref{gradient_evolution}  shows the metallicity gradients calculated for every bin with an orthogonal distance regression in both the [Fe/H]-R$_{\rm Gal}$ (blue circles) and [Fe/H]-r$_{\rm guid}$ (red circles) diagrams. We also compare these values with the gradient seen for Cepheids (green circle; \citealt{Genovali14}), which are young stars with a range in ages that is quite limited ($\sim$20-400 Myr), and the gradient-age relation observed by \citet{Casagrande11} on field stars (solid line). Based on our analysis, we are able to confirm the general behaviour that has been observed in the past: the [Fe/H]-R$_{\rm Gal}$ gradient changes with age and it is flatter for younger clusters. Interestingly, the gradients traced by the youngest groups are similar to those traced by Cepheids, and field stars. However, as we move towards older ages the cluster [Fe/H]-R$_{\rm Gal}$ gradient steepens, while the gradient observed from field stars goes in the opposite direction (see also \citealt{Anders17}). The behaviour traced by field stars is also in agreement with predictions by chemo-dynamical models of the Galactic disk, where radial migration is expected to smooth the gradient with time  (e.g., \citealt{Minchev18}).

We must therefore consider why is the age dependence of the gradient from open clusters is different to both observations and model predictions for field stars. In order to explain this paradox, we must conclude that either clusters migrate less than field stars or that the Galaxy, which destroys clusters living near gravitational potentials, is introducing a bias in their demographics. The first hypothesis seems unlikely, because the cluster masses are too small to significantly perturb spiral arms or giant molecular clouds and resist their impact \citep{Jilkova12,Gustafsson16}. On the other hand, based on the lack of old open clusters living near the Galactic mid-plane (Fig.~\ref{age_vz}), the second hypothesis is more plausible. In fact, on a first approximation, while stars can indiscriminately migrate inward and outward, the metal-rich clusters formed within the inner disk can only survive if they migrate outward where the Galactic potentials have a less destructive influence. On the other hand, metal rich clusters that do not migrate or migrate inward are quickly disrupted. The direct consequence of this Galactic selection effect is that the gradient traced by old clusters is steeper than the one seen for young clusters and field stars.

It is also interesting that the metallicity gradient traced in the [Fe/H]-r$_{\rm guid}$ diagram has a nearly constant dependence on age. In fact, the [Fe/H]-r$_{\rm guid}$ slope in the oldest bin lies in between the [Fe/H]-R$_{\rm Gal}$ gradient and the values observed for field stars. This observation suggests that the clusters' r$_{\rm guid}$ is less affected by radial mixing than R$_{\rm Gal}$. Therefore, r$_{\rm guid}$ is a better proxy of birth location for open clusters. This also implies that [Fe/H]-r$_{\rm guid}$ is a more fundamental space that can be used to study the chemical distribution of elements throughout the disk than the classical [Fe/H]-R$_{\rm Gal}$.

\subsection{The age dependence of chemical dispersion}

\begin{figure}
 \includegraphics[width=\columnwidth]{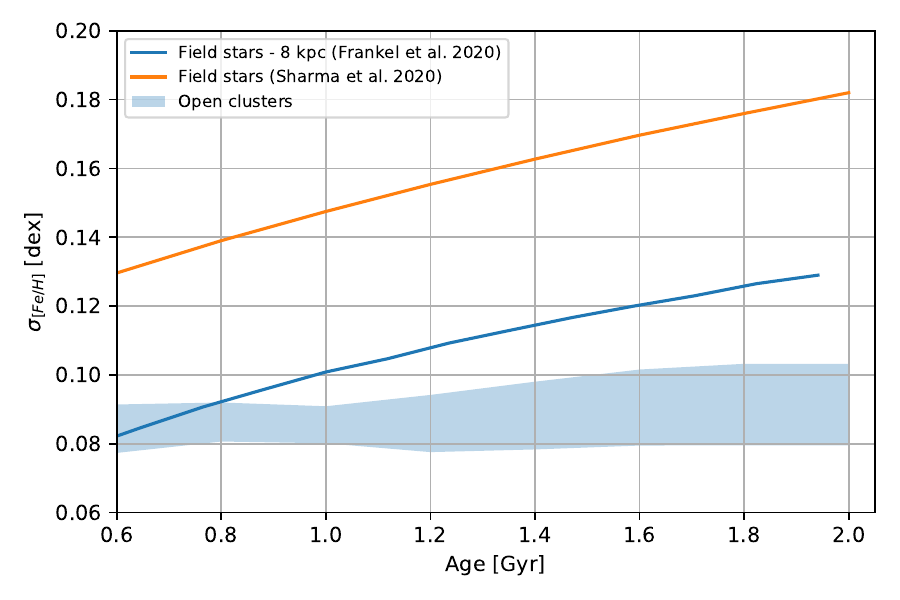}
 \caption{The figure shows the evolution of the chemical dispersion $\sigma_{\rm [Fe/H]}$ seen from open clusters (blue shaded area) and predicted from two chemo-dynamical models of the Galactic disk by \citet{Sharma20} and \citet{Frankel20}. }
 \label{chem_scatter_evolution}
\end{figure}

The chemical dispersion at a given location in the Galaxy is expected to increase with age because old stars had more time to migrate. In Fig.~\ref{chem_scatter_evolution}, we show the chemical dispersion $\sigma_{\rm [Fe/H]}$ of open clusters observed across the Galactic disk as a function of age (blue shaded area) and we compare it to the chemical dispersion predicted by the Galactic chemo-dynamical models by \citet{Sharma20} and \citet{Frankel20}. 

In order to obtain the observed $\sigma_{\rm [Fe/H]}$ among the open clusters, which is represented by the blue shaded area in Fig.~\ref{chem_scatter_evolution}, we first calculate for each clusters the residual between its observed [Fe/H] and the metallicity predicted by the relation [Fe/H]=$\alpha$$\times$R$_{\rm Gal}$+$\beta$, where $\alpha$ and $\beta$ are the mean values of the posteriors listed in Table~\ref{posteriors_tab}. Then, at a given age $\tau$, the blue area is limited by the standard deviation of the residuals obtained for clusters younger than $\tau$ (i.e., $\sigma_{\rm res(age<\tau)}$) and that from clusters older than $\tau$ (i.e., $\sigma_{\rm res(age\geq\tau)}$). Under the assumption that $\sigma_{\rm [Fe/H]}$ increases with cluster age, two values $\sigma_{\rm res(age<\tau)}$ and $\sigma_{\rm res(age\geq\tau)}$ are rough estimations of the lower and upper limits of the $\sigma_{\rm [Fe/H]}$ values at a given age $\tau$. Note that in this calculation we only include clusters with uncertainties $\Delta$Fe$<$0.1 dex and $\Delta$R$_{\rm Gal}<$0.5 kpc in order to limit the effect of very uncertain values on the residuals' distribution. 

The chemo-dynamical model from \citet{Sharma20} was calibrated over a sample of red-giants and main-sequence-turnoff stars observed by the GALAH and LAMOST \citep{Zhao12} surveys. They assumed a churning dispersion for a 12 Gyr old population equal to 1200  kpc km s$^{-1}$, which was taken from \citet{Sanders15}. On the other hand, \citet{Frankel20}, which used APOGEE red-clump stars and modelled the radial metallicity gradient in a slightly different way, obtained a churning dispersion equal to 875 kpc km s$^{-1}$ for a 12 Gyr old population. Also, note that the $\sigma_{\rm [Fe/H]}$ predicted by \citet{Sharma20} has no dependence from the Galactocentric radius, while the one from \citet{Frankel20} varies slightly with R$_{\rm Gal}$. Thus, for this latter model we plotted in Fig.~\ref{chem_scatter_evolution} the $\sigma_{\rm [Fe/H]}$ calculated at R$_{\rm Gal}$=8 kpc. As a result of the lower churning efficiency, the $\sigma_{\rm [Fe/H]}$ predicted by \citet{Frankel20} is systematically smaller than the one by \citet{Sharma20}. Not surprisingly, in both the models stellar migration increases $\sigma_{\rm [Fe/H]}$ with age.

In contrast, the $\sigma_{\rm [Fe/H]}$ evolution observed for open clusters shows two significative differences from the $\sigma_{\rm [Fe/H]}$ values predicted for field stars. First, the $\sigma_{\rm [Fe/H]}$ observed from clusters is considerably smaller than the values predicted by \citet{Sharma20}. The distribution of open clusters is also lower than the $\sigma_{\rm [Fe/H]}$ traced by the model of \citet{Frankel20}.  Secondly, the $\sigma_{\rm [Fe/H]}$ from clusters has an evolution that is nearly constant with age, in contrast with the increasing dispersion outlined by models.

These results provide a further indication that the surviving clusters have migrated less than what would be expected from stars of similar ages. Nevertheless, a comparison between observations and models is not fully straightforward. In fact, additional sources of chemical dispersion that likely affect our observations are neglected by models, such as the warp of the Galaxy and the effect of stellar activity on metallicity determination. In particular, this latter effect only affects the youngest clusters in our sample \citep{Galarza19,Baratella20}, increasing their chemical dispersion and making the chemical dispersion evolution appear flatter than what it actually is. However, stellar activity is supposed to affect only stars younger than 1Gyr \citep{Spina20}, thus it cannot explain a flat evolution of $\sigma_{\rm [Fe/H]}$ at greater ages.

\begin{figure*}
 \includegraphics[width=\textwidth]{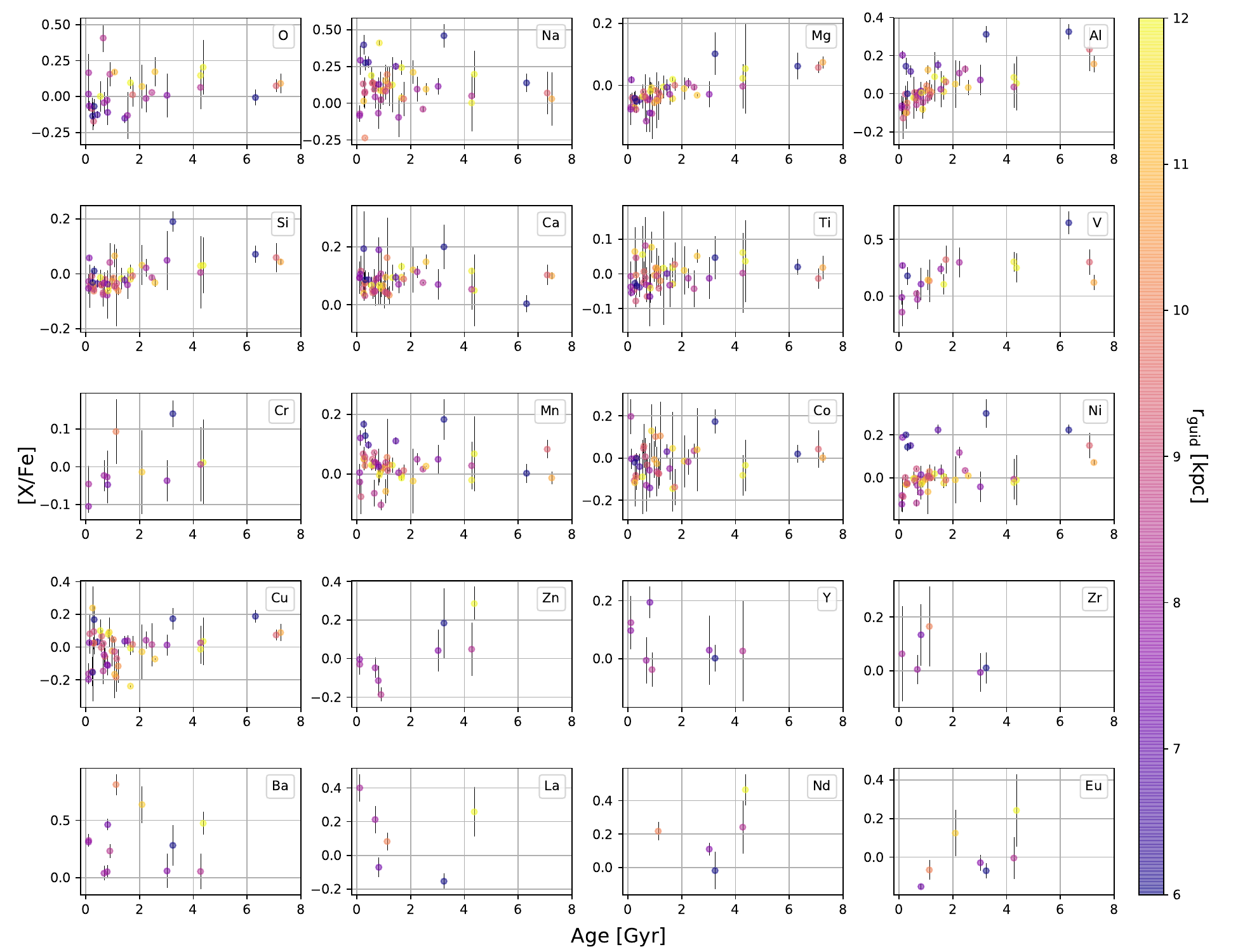}
 \caption{Abundance ratios [X/Fe] versus age. Open clusters are represented by circles colour coded as a function of their guiding radius r$_{\rm guid}$. The linear fit of the distributions traced by open clusters are shown by blue thick lines.}
 \label{XFe_age}
\end{figure*}

\bigskip

\section{Elements other than iron: galactic trends and chemical evolution}
\label{elements}
The variety of elements that can be detected in stellar atmospheres are produced from very different sites of nucleosynthesis (e.g., Type II or Type Ia supernovae, asymptotic giant branch stars) and have their own rates of release into the interstellar medium \citep{Kobayashi20}. Due to this diversity, the study of how abundances change in time and space throughout the Galactic disk is informative about element production channels and a myriad of other processes that are driving the history of the Milky Way (e.g., \citealt{Battistini16,Nissen15,Nissen20,Spina18,Lin20,Casali20}). 

In addition, a detailed study of the chemical composition of open clusters over a wide range of elements, ages, and Galactocentric radii also has key implications for ``chemical tagging'', a methodology based on the assumption that the chemical makeup of a star provides the fossil information of the environment in which it was formed. In fact, under this premise, it would be possible to tag stars that formed from the same material through clustering in the chemical space \citep{Freeman02,Bland-Hawthorn10,Ting15}. Regardless of the precision achievable in chemical abundances, the success of chemical tagging relies on the significance of two other critical factors: i) the level of chemical homogeneity of stars formed from the same giant molecular cloud; and ii) the chemical diversity of the interstellar medium in space and time.

Although theoretical studies have shown that clouds are chemically homogeneous within a 1 pc scale length \citep{Armillotta18} and that the typical chemical dispersion in the interstellar medium is statistically correlated on spatial scales of 0.5-1 kpc \citep{Krumholz18}, assessing in practice the efficacy of chemical tagging is not straightforward. In fact, chemical anomalies and variations are detected among members of the same association. However, these anomalies are either systematically ascribed to stars in different evolutionary stages (e.g., atomic diffusion; \citealt{Souto18,BertelliMotta18,Liu19}), or are rare and/or small compared to the typical uncertainties of large spectroscopic surveys (e.g., planet engulfment events; \citealt{Spina15,Spina18,Church20,Nagar20}). Therefore, despite a certain level of chemical inhomogeneity, recent studies have shown that it is still possible - in some measure - to distinguish cluster members from contaminants solely based on the chemical composition of stars (e.g., see \citealt{Blanco-Cuaresma18,Casamiquela20}). Given these promising premises, the detailed analysis of the chemical composition of open clusters is now necessary to provide model-independent constraints on the chemical diversity of the interstellar medium in space and time and to identify which are the elements that are mostly informative about age and birth location of stars.

Finally, knowledge on the chemical distribution of elements traced by open clusters is also useful to identify which are the species are the best tracers of stellar age or Galactocentric distance. For instance, the use of certain elements as chemical clocks could be affected by undesirable trends between abundances and Galactocentric distance (or metallicity; e.g., \citealt{Feltzing17,Casali20}).

\begin{figure*}
 \includegraphics[width=\textwidth]{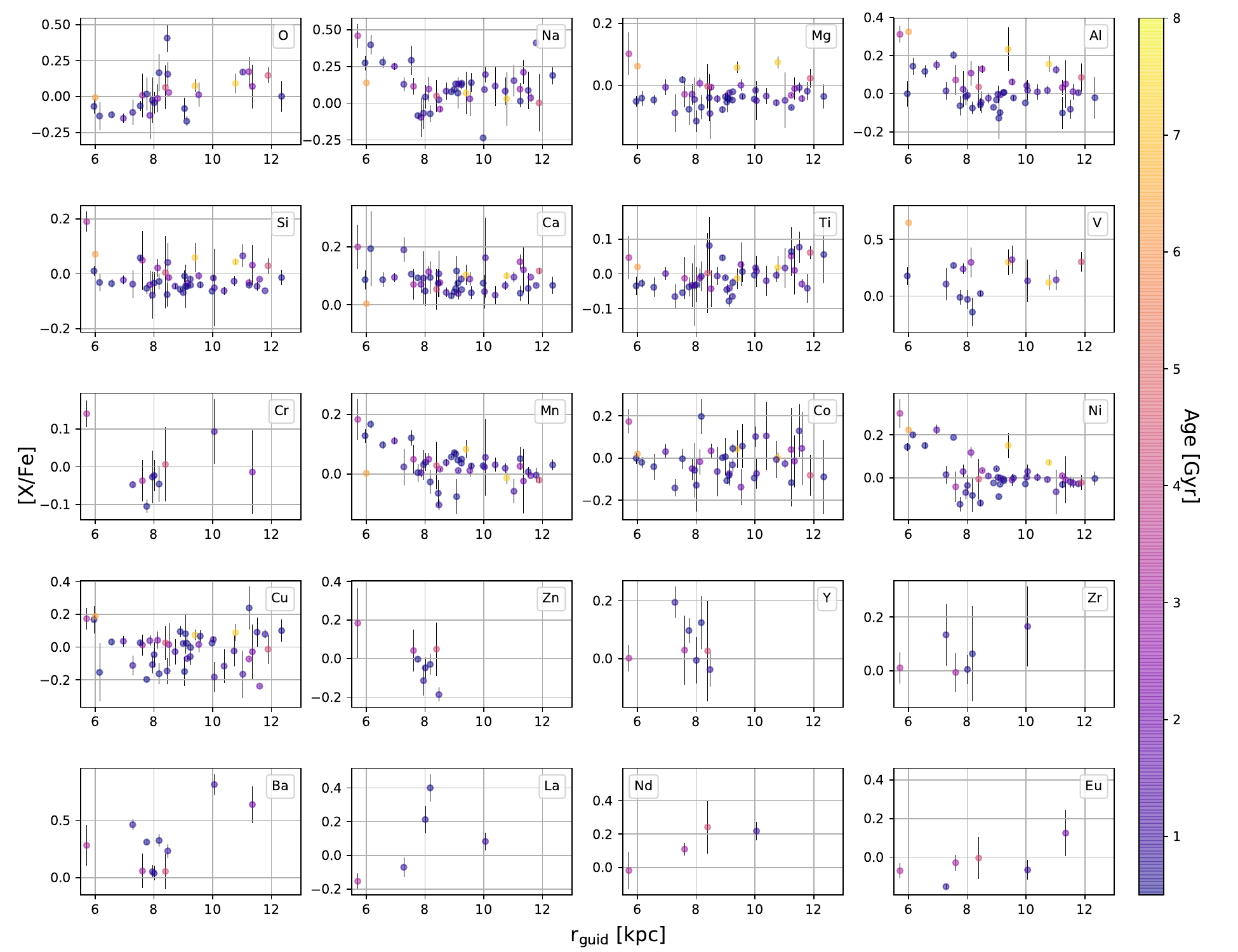}
 \caption{Abundance ratios [X/Fe] versus guiding radius r$_{\rm guid}$. Open clusters are represented by circles colour coded as a function of their age. The linear fit of the distributions traced by open clusters are shown by blue thick lines.}
 \label{XFe_rguid}
\end{figure*}

In Fig.~\ref{XFe_age} and \ref{XFe_rguid} we show the [X/Fe]-ages and [X/Fe]-r$_{\rm guid}$ distributions, respectively. In these diagrams we are plotting the same clusters shown in Fig.~\ref{FeH_dist} with the exclusion of those younger than 100 Myr and with uncertainties $\Delta$[X/Fe]$>$0.2 dex. We are considering 20 elements from oxygen to europium. These can be divided in four sub-classes: $\alpha$-elements (O, Mg, Si, Ca, and Ti), iron peak elements (V, Cr, Mn, Co, Ni, Cu, and Zn), odd-$z$ elements (Na, and Al), and neutron-capture elements (Y, Zr, Ba, La, Nd, and Eu). Only for the elements that are traced at least by 20 open clusters, we also perform a linear fit of their distributions. The resulting functions are plotted as blue thick lines, while the corresponding slopes with their uncertainties are listed in Table \ref{slopes}.

\setcounter{table}{4}
\begin{table}
\caption{[X/Fe]-age and [X/Fe]-r$_{\rm guid}$ slopes.}
\centering
\label{slopes}
{\small
\begin{tabular}{cccc}
\hline
Element & Class & [X/Fe]-age slope & [X/Fe]-r$_{\rm guid}$ slope \\
 &  & [10$^{2}$ Gyr$^{-1}$] & [10$^{2}$ kpc$^{-1}$] \\ \hline \hline
 O & $\alpha$ & 1.7$\pm$1.1 & 3.2$\pm$1.0 \\ 
 Na & odd-z &  -0.2$\pm$1.1 & -0.8$\pm$1.0  \\ 
 Mg & $\alpha$ &  1.9$\pm$0.2 & 0.2$\pm$0.3  \\ 
 Al & odd-z &  3.6$\pm$0.6 & -1.3$\pm$0.7  \\ 
 Si & $\alpha$ &  1.6$\pm$0.3 & -0.3$\pm$0.4  \\ 
 Ca & $\alpha$ &  0.0$\pm$0.4 & -0.3$\pm$0.3  \\ 
 Ti & $\alpha$ &  0.6$\pm$0.3 & 0.8$\pm$0.3  \\ 
 Mn & iron-peak &  0.0$\pm$0.5 & -1.2$\pm$0.4  \\ 
 Co & iron-peak &  0.7$\pm$0.7 & -0.7$\pm$0.7  \\ 
 Ni & iron-peak &  1.8$\pm$0.7 & -2.2$\pm$0.6  \\ 
 Cu & iron-peak &  1.7$\pm$0.9 & -0.3$\pm$0.8  \\ \hline \hline

\end{tabular}
}
\end{table}

\bigskip

\subsection{The $\alpha$-elements (O, Mg, Si, Ca, and Ti)}
Core collapse supernovae (SNe) are the evolutionary terminus of the most massive stars (M$\geq$10 M$_{\odot}$) and are the main site for the production of $\alpha$-elements. Owing to the short lifetime of massive stars ($\leq$10$^{-2}$ Gyr) these SNe pollute the interstellar medium essentially instantaneously compared to the timescale of secular Galactic evolution. The first core collapse SNe are also responsible for chemical enrichment in the very early stages of the Milky Way's history.

Mg, Si, and Ti in Fig.~\ref{XFe_age} share a similar behaviour characterised by a statistically significant decrease in their [X/Fe] abundance ratios going from old to young stellar ages: their slopes are larger than zero of more than two standard deviations. Also oxygen has a similar behaviour although its [X/Fe]-age slope is not larger two sigmas. This general trend closely resembles what is seen for nearby solar twins \citep{Nissen15,Spina16,Spina16b,Bedell18,Nissen20,Casali20} and it reflects the history of the Galactic disk, which formed from gas that was pre-enriched in $\alpha$-elements by the  thick disk and successively was enhanced in other metals by younger generations of stars. The flat distribution observed for Ca is due to the large contribution to this specie from Type Ia supernova which higher [Ca/Fe] for younger stars (e.g., see Fig. 2 in \citealt{Spina16b}): an opposite behaviour to that observed in other $\alpha$-elements, but that it is observed also in solar twins (e.g., \citealt{Spina16b,Casali20}). Another remarkable peculiarity of the [Ca/Fe] abundance ratios is that they consistently span super-solar values, around 0.10 dex. This is probably due to the fact that most of our clusters are younger than 4 Gyr, but it could also be due to a systematic due to the Solar Ca abundance chosen for GALAH DR3. However, this latter hypothesis is not corroborated by the abundance ratios measured in field stars. In fact, with the zero point chosen for GALAH DR3, the mean [Ca/Fe] ratio measured in stars of the solar vicinity is equal to 0.03$\pm$0.08 dex, while the typical difference in [Ca/Fe] between GALAH DR3 and APOGEE DR16 is equal to 0.07$\pm$0.12 dex \citep{Buder20}.

Only two elements, O and Ti, show a positive trend between in the [X/Fe] - r$_{\rm guid}$ diagrams (Fig.~\ref{XFe_rguid}). This is consistent with a scenario where the outflows rich of $\alpha$-elements produced by the tick disk have polluted the entire Galactic disk and, subsequently, the inner disk has been progressively enriched in other metals by Type Ia SNe \citep{Haywood13}. However, we do not observe similar trends for the other $\alpha$-elements, which is puzzling, especially for Mg, an element with a high contribution from core collapse SNe comparable to that of oxygen \citep{Kobayashi20}.

\subsection{The iron-peak elements (V, Cr, Mn, Co, Ni, Cu, and Zn)}
Type Ia supernova (SN Ia) explosions are thought to be the final fate of white dwarfs that undergo a thermonuclear explosion in a binary system; they are important producers of iron-peak elements. Owing to the relatively low mass of the SNe Ia progenitors, the timescales over which they operate are typically longer ($>$1 Gyr) than those of core collapse SNe. However, SNIa play an important role in astrophysics even though there are a variety of progenitors that are responsible for different SN Ia subclasses with diverging properties (\citealt{Kobayashi20}, and references therein).

Trends between [X/Fe] and age are observed in Fig.~\ref{XFe_age} for Ni and Cu. Although the slope observed for Cu is probably due to the high production rate of this element from hypernovae \citep{Kobayashi20}, Ni is mainly produced by SN Ia thus a positive slope, similar to that observed for $\alpha$-elements, was not expected. Furthermore, trends with Ni are not previously seen with open clusters \citep{Yong12,Donor20,Casamiquela20}. Therefore, even if the Ni slope is significative at the two-sigma level, it is not clear if it is real or it is mainly driven by the few oldest clusters. Instead, there are no trends with age for Mn and Co, which is consistent with a large contribution from SN Ia to these elements.

In the [X/Fe]-r$_{\rm guid}$ plane, negative slopes have been measured for Mn and Ni. While higher [Mn/Fe] ratios in the inner disk have been already seen by previous studies of open clusters \citep{Yong12,Magrini17,Donor20}, the same works have not observed similar trends for Ni. In any case, these trends observed by us indicate that Mn and Ni are largely produced by SN Ia, which is in perfect agreement with models of stellar nucleosynthesis (e.g., see Fig. 2 in \citealt{Spina16b}). Therefore, accordingly to these results, both the Mn and Ni abundance are strong indicators of stellar birth radii, similarly to the Fe abundance. In particular, Mn is probably the best indicator because [Mn/Fe] is not correlated with age.

Finally, the [X/Fe] slopes of V, Cr, and Zr have not been measured because these elements have been detected in less than 20 clusters. However, we notice that [V/Fe] probably has a positive correlation with age. However, a similar trend is seen neither in solar twins (e.g., \citealt{Bedell18,Casali20}), nor in the GALAH sample of field stars \citep{Lin20}.

\subsection{The odd-z elements (Na, and Al)}

Even though Na is mainly produced by core collapse SNe, with contribution percentages similar to those of O, Mg, and Si (\citealt{Kobayashi20}, and references therein), there is no trend between [Na/Fe] and age. However, the peculiarity of this element, in comparison to the other $\alpha$-elements, is that its synthesis is controlled by the neutron
excess \citep{Timmes95}, meaning that the Na mass produced in SNe is highly dependent from the stellar metallicity (see Fig. 5 in \citealt{Kobayashi06}). Thus, the oldest generations of stars, those with the lowest metallicity, did not deliver to the interstellar medium as much Na as the later core collapse SNe progenitor did. As a result, the distribution of clusters in the [Na/Fe]-age diagram is flat. Similarly, also the [Na/Fe]-r$_{\rm guid}$ diagram shows a flat distribution which is consistent to other $\alpha$-elements such as Mg and to what is previously observed by \citep{Yong12}.

Conversely to Na, the [Al/Fe]-age diagram shows a neat positive trend, resembling the behaviour of $\alpha$-elements. In fact, accordingly to models of stellar nucleosynthesis, Al has a contribution from core collapse SNe similar to that of O and Mg. Similarities in the nucleosynthetic evolution of Al and $\alpha$-elements are also observed in solar twins (e.g., \citealt{Bedell18,Casali20}) and previous studies of open clusters (e.g., \citealt{Yong12}). It is also interesting to point out that, among all the elements analysed here, [Al/Fe] shows the steepest correlation with age. For this reason, the Al element is commonly used in chemical clocks \citep{Spina18,Casali20}. However, [Al/Fe] also shows a correlation with r$_{\rm guid}$ which is undesirable for reliable age indicators.

\subsection{The neutron-capture elements (Y, Zr, Ba, La, Nd, and Eu)}

The neutron-capture elements can be produced through the slow- (s-) process, which occurs in low- and intermediate-mass stars (i.e., 1-8 M$_{\odot}$) during their AGB phase (e.g., \citealt{Karakas16}), or through a rapid- (r-) process. While the main production channel for the r-process is still debated, neutron star mergers have in recent years been verified as a viable channel. Other production sites including SN explosions, black hole/neutron star mergers and electron capture supernovae \citep{Argast04,Cowan04,Surman08,Thielemann11,Korobkin12}. Following the contribution percentages taken from \citet{Bisterzo14}, more than 70$\%$ of Y, Ba, and La is produced through the s-process, around the 60$\%$ of Zr and Nd is produced from the s-process, while most of Eu ($>$90$\%$) is produced through the r-process.

Unfortunately, the neutron-capture elements are observed in a small number of clusters, therefore their [X/Fe] slopes have not been measured. Nevertheless, Fig.~\ref{XFe_age} shows  that most of the youngest clusters have super-solar abundances of the elements produced through the s-process (i.e., Y, Zr, Ba, and La). This is expected from a high production rate of these elements from AGB stars and it is also confirmed by observations of both solar twins (e.g., \citealt{Spina18,Casali20}) and open clusters \citep{Magrini18}. However, note that \citet{DOrazi12,DOrazi17} found young clusters having super-solar abundances of Ba, but not of the other s-process elements. This may be because the Ba lines used in the spectroscopic analysis are particularly sensitive to stellar activity. In fact, the Ba lines that are commonly used to derive its abundance in stars form in the upper atmosphere, where strong magnetic fields may be present and, hence, they may be incorrectly modelled \citep{Reddy17,Spina20}.

\section{Conclusions}
\label{conclusions}
Open clusters are unique tracers of the history of our own Galaxy. In this work, we used astrometric information from \textit{Gaia} to identify the stellar members of the 226 open clusters that potentially fall in the footprints of APOGEE or GALAH surveys (see Section~\ref{Membership}). Based on our membership analysis, members of 205 clusters were actually observed by these two surveys. In Section~\ref{properties}, we calculated the physical and kinematic properties of these latter. Among these 205 clusters, 134 have stars with abundance determinations and that satisfy the selection criteria listed in Section~\ref{Chemical_abundances}. We have used this invaluable information to study the chemical distribution of elements throughout the Galactic disk. For instance, we find that our sample of open clusters outlines a radial metallicity gradient in the [Fe/H]-R$_{\rm Gal}$ diagram of $-$0.076$\pm$0.009 dex kpc$^{-1}$, which is in agreement with previous studies based on open clusters \citep{Jacobson16,Carrera19,Donor20}.

The most relevant result of our study comes from combined kinematic and chemical observations providing additional insights on how the Galaxy is shaping the clusters' demographic. Namely, in Sections \ref{gradient} and \ref{survivors} we observe that 
\begin{itemize}
\item The chemical dispersion of open clusters in the [Fe/H]-r$_{\rm guid}$ plane is smaller than that measured in [Fe/H]-R$_{\rm Gal}$ diagram (Fig. \ref{FeH_dist} and \ref{all_posteriors}).
\item The mid-plane of the inner Galaxy lacks of old clusters, which, instead, can be found at high altitudes from the plane (Fig.~\ref{age_vz}).
\item Gradient traced by old clusters in the [Fe/H]-R$_{\rm Gal}$ diagram is steeper than that seen for young clusters and cepheids. This observation is different to what is outlined by field stars (Fig.~\ref{gradient_evolution}).
\item The R$_{\rm Gal}$  and r$_{\rm guid}$ metallicity gradients have different age dependencies (Fig.~\ref{gradient_evolution}).
\item The chemical dispersion traced by clusters is smaller than that observed through field stars of similar ages (Fig.~\ref{chem_scatter_evolution}).
\end{itemize}

All these pieces of evidence, for the first time assembled together in a self-consistent study of open clusters provide new clues on how clusters live and die in our Galaxy. For example, it is now clear that in-plane potentials (e.g., spirals, bars, molecular clouds) are more perilous to clusters than passages through the disk. In fact, while these latter would induce a lack of high-altitude clusters \citep{Martinez-Medina17}, we observe that the Galaxy is preserving the associations that are living far from the plane and is quickly destroying all the others. As a consequence of this selection effect imposed by the Milky Way, open clusters trace the distribution of elements across the Galactic disk differently from field stars. This happens because the clusters that we are observing today are either the youngest or those living far from the gravitational potentials. In any case, these are the clusters for which we expect a minimal degree of churning compared to what typical field stars have experienced.

Furthermore, the observations described above for the first time show that, conversely to the case of field stars, L$_{\rm z}$ (or r$_{\rm guid}$) is a more fundamental parameter than R$_{\rm Gal}$ for open clusters against which the chemical distribution of elements should be studied.

A detailed study of the age distribution of open clusters is extremely important to improve our understandings on how they form and are disrupted in our Galaxy. It is very likely that not all open clusters dissipate on similar timescales. On the contrary, a complex - but interesting - interplay between their spatial distribution, ages, kinematics, masses and densities would probably explain the proportion between open cluster formation and their survival rates. The study of this interplay is beyond the scope of our current project. Nevertheless, the evidence that the existing open clusters are not redistributed across the Galaxy like field stars is a key factor for understanding how these two populations trace in different ways the chemical enrichment of the Milky Ways.

This has extremely important implications for future efforts to disentangle the chemical distribution of elements in our Galaxy from radial migration. For instance, our results indicate how past or future efforts of modelling the dispersion of clusters across the Galactic disk (e.g., \citealt{Quillen18}) are incomplete when they neglect the impact of cluster dissipation. Furthermore, it is now evident that chemo-dynamical models of our Galaxy trained and tested on field stars cannot be indiscriminately used to study the demographic of open clusters, and \textit{vice versa}. On the contrary, field stars and open clusters are two faces of the same coin, with their similarities and differences. Therefore models aiming at a distinct parameterisation of these two tracers would likely unlock deeper and more comprehensive insights on our Galaxy than whats has been done so far. 

For instance, in a pioneering work, \citet{Minchev18} proposed a method to infer the stellar birth radii based on stellar age and metallicity with the help of an assumed interstellar medium evolution profile. A similar idea was also exposed by \citet{Ness19}. More recently, the idea was borrowed by \citet{Feltzing20} and applied to a sample of stars observed by the APOGEE survey in order to investigate the relative importance of blurring and churning. Similar studies were also carried out by \citet{Frankel18,Frankel19,Frankel20}. Unfortunately all these approaches rely entirely on models of the interstellar medium chemical evolution, thus their outputs cannot be used to constrain different scenarios of the formation and evolution of the Galactic disk, e.g., the inner/outer disks dichotomy \citep{Haywood13,Snaith15}, the double accretion scenario \citealt{Chiappini97,Spitoni19a}, the central stellar burst \citep{Grand18}, or merger scenarios \citep{Calura09,Buck20}. Therefore, the formation of the Galactic disk is still an unsettled question and it is not trivial to distinguish between the different scenarios without making assumptions on stellar migration or on the shape of the interstellar medium evolution profile. Instead, the possibility that an existing cluster is less likely to have churned than field stars would justify the use of clusters as benchmarks for the chemical composition of the gas at their present r$_{\rm guid}$ and as the best model-independent tracers of the role of radial migration in the Galactic disk, as it has recently been experimented by \citet{Chen20}. 

Finally, the spatial and temporal evolution of abundance ratios [X/Fe] studied in Section~\ref{elements}. Our analysis (see Fig. \ref{XFe_age} and \ref{XFe_rguid}) has revealed further insights into the nucleosynthesis' sites and timescales of a large variety of elements (O, Na, Mg, Al, Si, Ca, Ti, V, Cr, Mn, Co, Ni, Cu, Zn, Y, Zr, Ba, La, Nd, and Eu).

\section*{Acknowledgements}
This work is based on data acquired through the Australian Astronomical Observatory, under programmes: A/2019A/01 (Hierarchical star formation in Ori OB1), A/2014A/25, A/2015A/19, A2017A/18 (The GALAH survey); A/2015A/03, A/2015B/19, A/2016A/22, A/2016B/12, A/2017A/14 (The K2-HERMES K2-follow-up program); A/2016B/10 (The HERMES-TESS program); A/2015B/01 (Accurate physical parameters of Kepler K2 planet search targets); S/2015A/012 (Planets in clusters with K2). We acknowledge the traditional owners of the land on which the AAT stands, the Gamilaraay people, and pay our respects to elders past and present. LS and AIK acknowledge financial support from the Australian Research Council (discovery Project 170100521) and from the Australian Research Council Centre of Excellence for All Sky Astrophysics in 3 Dimensions (ASTRO 3D), through project number CE170100013. Y.S.T. is grateful to be supported by the NASA Hubble Fellowship grant HST-HF2-51425.001 awarded by the Space Telescope Science Institute. TCG acknowledges support by the Spanish Ministry of Science, Innovation and University (MICIU/FEDER, UE) through grants RTI2018-095076-B-C21 and the Institute of Cosmos Sciences University of Barcelona (ICCUB, Unidad de Excelencia \textit{Mar\'{\i}a de Maeztu}) through grant CEX2019-000918-M. A.~R.~C. is supported in part by the Australian Research Council through a Discovery Early Career Researcher Award (DE190100656). SLM and DZ acknowledge the support of the Australian Research Council through grant DP180101791. MA has been supported by the Australian Research Council (projects FL110100012 and DP150100250). TTG acknowledges financial support from the Australian Research Council (ARC) through an Australian Laureate Fellowship awarded to JBH. 

\section*{Data Availability}
The data underlying this article are available in the article and in its online supplementary material at the CDS.


\setcounter{table}{0}
\begin{table*}
\centering
\caption{Cluster membership probabilities analysis and  cross-match with GALAH and APOGEE datasets - full table available online at the CDS}
\label{members}
\medskip
\begin{tabular}{ccccccccc}
\hline
R.A. (J2050) & DEC (J2050) & source\_id & sobject\_id & APOGEE\_ID & Cluster & P\_mem & P\_mem\_CG20 & ... \\
~[deg] & [deg] & & & & & & & ... \\ \hline \hline
80.699829 & 1.716740 & 3234192360520220928 & 151227004202188 & 2M05224795+0143002 & ASCC\_16 & 0.82 & 1.0 & ... \\
80.454788 & 0.587694 & 3221677478654607360 & 190211002201218 & 2M05214914+0035156 & ASCC\_16 & 0.14 & 0.0 & ... \\
80.878777 & 1.855169 & 3234289048824132480 & --- & 2M05233090+0151185 & ASCC\_16 & 0.98 & 0.0 & ... \\
81.033539 & 2.463022 & 3234348869128542080 & --- & 2M05240804+0227468 & ASCC\_16 & 0.73 & 1.0 & ... \\
80.841110 & 1.976155 &  3234295130497815424 & 190212002001261 & --- & ASCC\_16 & 0.46 & 0.0 & ... \\
... & ... & ... & ... & ... & ... & ... & ... & ... \\ \hline
\hline
\end{tabular}
\end{table*}

\setcounter{table}{1}
\begin{table*}
\centering
\caption{Physical and kinematic properties of open clusters - full table available online at the CDS}
\label{cluster_kin}
\medskip
\begin{tabular}{cccccccccc}
\hline
Cluster & log$_{\rm10}$ Age/Gyr & N\_members & RA & RA\_std & DEC & DEC\_std & plx & plx\_std & ... \\
 & & & [deg] & [deg] & [deg] & [deg] & [mas] & [mas] & ... \\ \hline \hline
 ASCC\_16 & 7.13 & 289 & 81.171005 & 0.32 & 1.652395 & 0.37 & 2.83 & 0.14 &    ... \\
 ASCC\_19 & 7.02 & 264 & 82.046997 & 0.65 & -1.976446 & 0.51 & 2.77 & 0.11 &    ... \\
 ASCC\_90 & 8.91 & 67 & 264.756836 & 0.46 & -34.892029 & 0.53 & 1.72 & 0.05 &  ... \\
 ASCC\_99 & 8.55 & 82 & 282.218628 & 0.54 & -18.527493 & 0.66 & 3.38 & 0.12 &    ... \\
 Alessi\_24 & 7.86 & 152 & 260.772339 & 0.94 & -62.673843 & 0.56 & 2.06 & 0.07 &   ... \\
... & ... & ... & ... & ... & ... & ... & ... & ... & ... \\ \hline
\hline
\end{tabular}
\end{table*}

\setcounter{table}{2}
\begin{table*}
\centering
\caption{Chemical properties of open clusters - full table available online at the CDS}
\label{cluster_chem}
\medskip
\begin{tabular}{cccccccccccccc}
\hline
Cluster & Fe & Fe\_err & N\_Fe & O & O\_err & N\_O & Na & Na\_err & N\_Na & Mg & Mg\_err & N\_Mg & ... \\ \hline \hline
ASCC\_16 & -0.11 &  0.04 & 14 &  0.23 &  0.06 &    4 &  0.14 &  0.09 &     4 & -0.09 &  0.07 &     5 & ... \\
ASCC\_19 & -0.05 &  0.05 & 20 &  0.38 &  0.04 &    5 & -0.05 &  0.04 &     1 & -0.05 &  0.07 &    10 & ... \\
Alessi\_24 &  0.02 &  0.03 & 2 & --- & --- &    0 & --- & --- &     0 & --- & --- & 0 &  ... \\
Alessi\_9 & -0.04 &  0.06 & 3 &  0.23 &  0.15 &    2 & -0.03 &  0.04 &     1 &  0.02 &  0.13 &     2 & ... \\
BH\_211 &  0.214 &  0.015 & 3 & -0.18 &  0.04 &    3 &  0.25 &  0.02 &     3 & -0.02 &  0.02 &     3 &   ... \\
... & ... & ... & ... & ... & ... & ... & ... & ... & ... & ... & ... & ... & ... \\
\hline \hline
\end{tabular}
\end{table*}



\bibliographystyle{mnras}
\bibliography{Bibliography} 



\bsp	
\label{lastpage}
\end{document}